\documentclass[sigconf]{acmart}
\usepackage{url}
\usepackage{amsmath,amssymb,amsfonts}
\usepackage{accents}
\usepackage[linesnumbered,ruled]{algorithm2e}
\usepackage[noend]{algpseudocode}
\usepackage{graphicx}
\usepackage{textcomp}
\usepackage{xcolor}
\usepackage{comment}
\usepackage{multicol,multirow}
\usepackage[T1]{fontenc}
\usepackage{todonotes}
\usepackage{caption}
\usepackage{wrapfig}
\usepackage{subfig}
\newsavebox{\measurebox}
\usepackage{threeparttable}
\usepackage[font=small]{caption}
\usepackage{enumitem}
\usepackage{array}
\usepackage{makecell}
\usepackage{pbox}
\usepackage{stfloats}

\usepackage{hyperref}
\hypersetup{
    colorlinks=false
}

\makeatletter
\newcommand{\algorithmfootnote}[2][\footnotesize]{%
  \let\old@algocf@finish\@algocf@finish% Store algorithm finish macro
  \def\@algocf@finish{\old@algocf@finish% Update finish macro to insert "footnote"
    \leavevmode\rlap{\begin{minipage}{\linewidth}
    #1#2
    \end{minipage}}%
  }%
}
\makeatother

\AtBeginDocument{%
  \providecommand\BibTeX{{%
    \normalfont B\kern-0.5em{\scshape i\kern-0.25em b}\kern-0.8em\TeX}}}

\copyrightyear{2022}
\acmYear{2022}
\setcopyright{rightsretained}
\acmConference[CHASE' 22]{ACM/IEEE International Conference on Connected Health: Applications, Systems and Engineering Technologies}{November 17--19, 2022}{Washington, DC, USA}
\acmBooktitle{ACM/IEEE International Conference on Connected Health: Applications, Systems and Engineering Technologies (CHASE' 22), November 17--19, 2022, Washington, DC, USA}
\acmDOI{10.1145/3551455.3559151}
\acmISBN{978-1-4503-9476-5/22/11}

\begin{document}

% \setlength{\abovedisplayskip}{3pt}
% \setlength{\belowdisplayskip}{3pt}

\title{Design and Validation of an Open-Source Closed-Loop Testbed for Artificial Pancreas Systems}

% \author{Regular Paper}
% \begin{comment}

% \email{}
% \orcid{}
\author{Xugui Zhou}
\email{xugui@virginia.edu}
\affiliation{%
  \institution{University of Virginia}
  \city{Charlottesville}
  \state{Virginia}
  \country{USA}
  \postcode{22904}
}

\author{Maxfield Kouzel}
\email{mak3zaa@virginia.edu}
\affiliation{%
  \institution{University of Virginia}
  \city{Charlottesville}
  \state{Virginia}
  \country{USA}
  \postcode{22904}
}

\author{Haotian Ren}
\email{hr3xw@virginia.edu}
\affiliation{%
  \institution{University of Virginia}
  \city{Charlottesville}
  \state{Virginia}
  \country{USA}
  \postcode{22904}
}

\author{Homa Alemzadeh}
\email{ha4d@virginia.edu}
\affiliation{%
  \institution{University of Virginia}
  \city{Charlottesville}
  \state{Virginia}
  \country{USA}
  \postcode{22904}
}
% \end{comment}

\renewcommand{\shortauthors}{X. Zhou, M. Kouzel, H. Ren, H. Alemzadeh.}

\begin{abstract}
The development of fully autonomous artificial pancreas systems (APS) that independently regulate the glucose levels of patients with Type 1 diabetes has been a long-standing goal of diabetes research. A significant barrier to progress is the difficulty of testing new control algorithms and safety features, since clinical trials are time- and resource-intensive. To facilitate ease of validation, we propose an open-source APS testbed that can integrate state-of-the-art APS controllers and glucose simulators with a novel fault injection engine. The testbed is used to reproduce the blood glucose trajectories of real patients from a clinical trial conducted over six months. We evaluate the performance of two closed-loop control algorithms (OpenAPS and Basal Bolus) using the testbed and find that these control algorithms are able to keep blood glucose in a safe region 93.49\% and 79.46\% of the time on average, compared with 66.18\% of the time for the clinical trial. The fault injection engine simulates the real recalls and adverse events reported to the U.S. Food and Drug Administration (FDA) and demonstrates the resilience of the controller in hazardous conditions. We use the testbed to generate 25 years of synthetic data representing 20 different patient profiles with realistic adverse event scenarios, which would have been expensive and risky to collect in a clinical trial. %The proposed testbed is a valid tool that can be used by the research community to demonstrate the effectiveness of different control algorithms and safety features for APS.
\end{abstract}


%\ccsdesc[300]{Computer systems organization~Medical Cyber-Physical Systems}


\settopmatter{printfolios=true}
%%lewis2020yourself
\maketitle

\vspace{-1em}
\section{Introduction}
% MCPS

% Virtual patient model for MCPS
% Cite the Virtual heart paper, Philip, surgical robot 
% Requirement have a realistic testbed that does not harm the patients
% For aps we could find anything in the literature, so we design and develop this work

% Enables integrating safety feature into the 

% \vspace{-0.5em}
% P1: MCPS are safety critical and the importance of safety assurance. Developing high fidelity simulators that can capture the variety of patient profiles and physiological dynamics and changes as the result of changes in environment and activity is very important. Another important need is the ability to simulate the unexpected events such as accidental faults, human errors, and attacks that lead to adverse events. Such closed-loop testbeds can enable verification of control algorithms and safety features before the additional cost of clinical trials and reducing the possibility of harm to patients. 

% P1 --- importance
Medical cyber-physical systems (MCPS) apply computational algorithms to regulate and control complex processes in the human body. By nature, these are safety-critical systems that directly affect the health of those who use them, so safety assurance is crucial. Developing high-fidelity simulators that can capture a variety of patient profiles and physiological dynamics as well as react to changes in environment and activity is very important, as it allows considerable research on the safety assurance of MCPS to be performed at an accelerated rate while avoiding unnecessary risks to patients. 
Another essential need is the ability to simulate unexpected events such as accidental faults, human errors, and attacks that lead to adverse events. Such closed-loop testbeds can enable verification of control algorithms and safety features before the additional cost of clinical trials and reduce the possibility of harm to patients.

% The use of simulators is vital in the development of healthcare technologies, which allows significant research to be performed at an accelerated rate while circumventing unnecessary risks to the patient and costs related to animal or clinical testing [6]. Simulators have played a prominent role in the development of many important areas of biomedical research, such as anesthesia administration [7], HIV therapy [8], minimal...

% P2 --- 
% P2: State of the art work on closed-loop testbeds with virtual patient models for verification. For example, UPenn, MdPnP, Philip's, Surgical robot
% UPenn - heart model for pacemaker validatin
% MDPnP - treatment validation protocol; cardiac arrest case study
% Pulse Physiology Engine - uses lumped-parameter model (i.e. circuit analog for physiological systems) to simulate the cardiovascular system and pharmokinetic interactions
% Surgical robot (my characterization of this work may not be entirely correct, might need a different paper)
Progress has been made in developing these closed-loop testbeds in many vital areas of MCPS, such as a virtual heart model for pacemaker verification~\cite{zhihao_jiang_using_2010}, simulation platforms for robot-assisted surgery \cite{alemzadeh2015software, homa2016targeted}, and joint simulation of the cardiovascular system and pharmacokinetic interactions \cite{bray_pulse_2019, gessa2019towards}.
Further, validation for not just medical controllers but also medical protocols for treatment such as cardiac arrest \cite{wu_treatment_2014} has proven effective at verifying experimental treatments before it ever reaches actual patients. 
However, few closed-loop testbeds for artificial pancreas systems (APS) are found in the literature \cite{toffanin2020silico,schmitzer2022efficient}, not considering variety of patient profiles and adverse event scenarios.
%with limited virtual patients. 

% . A verification testbed for pacemaker software was developed using a model of the heart \cite{zhihao_jiang_using_2010}, and advanced models of surgical robots allow for early detection of hazardous runtime errors during surgery \cite{li_runtime_2022}. Pulse, a more general modelling software engine \cite{bray_pulse_2019}, has also been shown to accurately simulate the cardiovascular system and pharmokinetic interactions. 

% P3: APS is one of the good examples of MCPS which is approved by FDA through clinical trials. although lots of work in the literature in diabetes on glucose simulators and controllers and clinical trials, no closed-loop testbeds and not incorporating the safety mechanisms or simulation of adverse events.
% APS is one of the good examples of MCPS which is approved by FDA through clinical trials. The UVA-Padova Type 1 Diabetes Simulator (UVA-Padova) was 
% P3
% simulators - Glucosym, T1DS, others?
% control algorithm papers?
% studies - PSO3, DCLP, others?
The APS is a good example of a promising MCPS that the U.S. Food and Drug Administration (FDA) approved through clinical trials. 
Much work has been done in the literature to develop realistic diabetes simulators \cite{man_uvapadova_2014,visentin2018uva,KanderianT1}, % or in-silico trials \cite{toffanin2020silico,schmitzer2022efficient}, 
design advanced control algorithms to maintain glucose concentration at healthy levels \cite{lewis2020yourself, tandemAlgorithm}, and conduct large-scale clinical trials \cite{brown2019six, breton_randomized_2020, maahs_randomized_2014}. However, to the best of our knowledge, none of these works considered closed-loop integration of simulators, controllers, and safety mechanisms or the simulation of adverse events. 
Control software must be tested in a broad spectrum of environments and with a variety of patient profiles and physiological dynamics in order to be fully verified.
However, it is challenging to identify and tune simulator parameters that characterize the patient profiles and physiological dynamics (e.g., meals and physical activity) and generate realistic and representative adverse event scenarios.
% Multiple powerful simulators have been developed for the APS \cite{man_uvapadova_2014,visentin2018uva,KanderianT1}, much work has been done to develop advanced control algorithms \cite{?} to maintain glucose concentration at healthy levels, and several large-scale clinical trials have been carried out \cite{?}, yet there are no closed-loop testbeds for APS that incorporate safety mechanisms or simulation of adverse events. Control software must be tested in a broad spectrum of environments in order to be fully verified.

% P4
% P4: This paper presents the design and validation of an open-source closed-loop testbed for APS by ....

To fill this gap, this paper presents the design and validation of an open-source, closed-loop testbed for APS that can integrate state-of-the-art Type 1 diabetes glucose simulators (e.g., Glucosym \cite{Glucosym} or UVA-Padova \cite{man_uvapadova_2014}) and insulin delivery control software (e.g., OpenAPS or Basal-Bolus) together with an adverse events simulator that simulates the typical adverse events reported to the FDA. %to help with the resilience testing and quality verification of APS devices. 
We assess the validity of the proposed testbed by comparing the simulator outputs, controller outputs, and closed-loop outcomes with the data collected from a clinical trial. 
An optimization method is also proposed to reconstruct the blood glucose (BG) traces from a real-world clinical trial by estimating the patient profiles.

Experimental results show that the integrated glucose simulators can well reproduce the BG traces in the clinical trial, given the exact insulin dosages in the trial, and the integrated (open-loop) controllers keep a low mean squared error with the actual pump outputs in the clinical trial. The closed-loop simulations can keep blood glucose in a safe region 93.49\% and 79.46\% of the time on average, compared with 66.18\% of the time for the clinical trial.
Because the testbed aims to provide a platform to validate different control algorithms and safety features efficiently, we provide an easy-to-use fault injection implementation for both simulators and detail how the simulated faults compare to real fault scenarios in the device recalls and adverse events reported to the FDA.

The main contributions of the paper are as follows:
\begin{itemize}[leftmargin=*]

\item Proposing an open-source closed-loop platform by integrating classical artificial pancreas (AP) controllers (OpenAPS and Basal-Bolus) and glucose simulators (Glucosym and UVA-Padova), which 
(1) enables experimental evaluation of different controllers, prediction algorithms, and pump functionalities, or other studies on APS that do not have access to clinical patients; 
(2) integrates a fault injection engine to simulate potential safety and security issues and adverse events in APS caused by accidental software and hardware faults, human errors, or malicious attacks.

%\item Developing an open-source testbed that allows research on developing APS and control algorithms that does not have access to clinic patients to be performed at an accelerated rate while avoiding the unnecessary risks in a clinical test.

\item Developing an optimization method to estimate patient profiles and reconstruct BG trajectories from clinical studies in simulation, which extends the number of virtual patients in the testbed and enables the assessment of different algorithms and adverse event scenarios using existing patient datasets without the need for doing new clinical trials for each algorithm.

\item Simulating example adverse event scenarios involving APS based on the data reported to the FDA, resulting in the generation of 25 years of synthetic data with different adverse event scenarios for 20 different patient profiles, which would have been very expensive and risky to collect from real patients or clinical trials.

\item Assessing the validity of the APS testbed based on a publicly-available clinical trial dataset that includes six months of data for 168 diabetic patients. The results show that the APS simulators and controllers can reasonably approximate the actual glucose monitor and insulin pump functionalities with an average mean squared error of $3.97x10^{-3}$. Also, the closed-loop simulations show that the integrated OpenAPS and Basal-Bolus controllers can keep the blood glucose in the target range for at least 11.28\% longer time than the controllers used in clinical trials. 

\end{itemize}

%=============================================================%
\section{Background}
\begin{figure}[t]
    \centering
    \includegraphics[width=0.55\columnwidth]{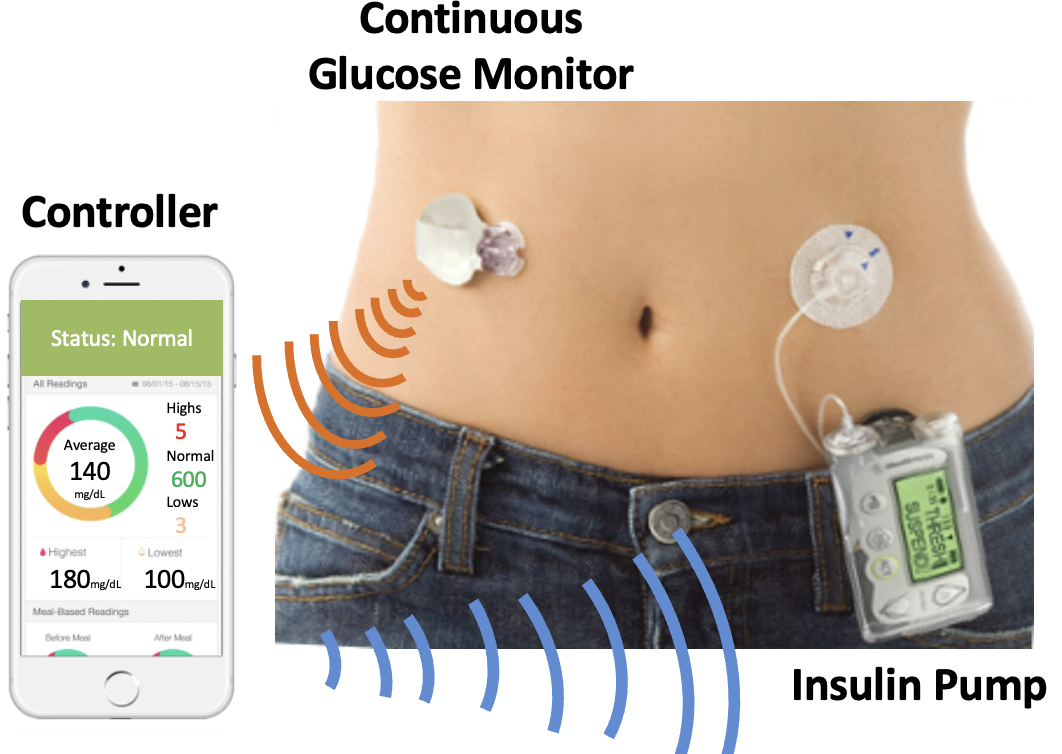}
    \vspace{-1em}
    \caption{Artificial Pancreas System}
    \label{fig:controller}
    \vspace{-2em}
\end{figure}

\textbf{Artificial Pancreas System (APS):} APS is a medical control system that regulates the glucose levels of people with Type 1 diabetes \cite{atkinson2014type}, who cannot regulate their own glucose levels because their pancreas does not produce insulin on its own.
%for whom the pancreas does not produce insulin on its own, so they are unable to regulate their glucose by themselves.
An APS is responsible for regulating Blood Glucose (BG) dynamics by monitoring BG concentration in the patient's body through sensor data collected from a Continuous Glucose Monitor (CGM) and by providing insulin at the correct rate to the patient through an insulin pump. The control software estimates the current patient status (e.g., glucose value, insulin on board (IOB)) and calculates the next recommended insulin rate value to be delivered to the patient. The typical APS is shown in Fig.~\ref{fig:controller}.

The development of a fully autonomous artificial pancreas system has been a long-standing goal of research into Type 1 diabetes. A milestone was reached when the first commercially available closed-loop APS that automated basal insulin delivery was approved by the FDA in 2017 \cite{saunders_minimed_2019}. Nevertheless, it still required user input to indicate insulin boluses. An ideal controller must be able to account for unannounced meals and variety of physical activities. While there are existing algorithms for the detection of meals and different activities~\cite{samadi_automatic_2018}, they have yet to be incorporated into APS controllers.

% APS simulation
% fundamentally built upon a patient model that accurately reflects how the body reacts to insulin dosage delivered by the pump and chosen by the controller.
\textbf{APS Simulation:}
The APS controllers and glucose simulators are fundamentally built on the patient models that reflect how the body reacts to insulin dosage. Two main patient models available in the literature that we use in this paper are Glucosym and UVA-Padova. These simulators use systems of differential equations to model the change of state variables important to glucose and insulin kinetics. The equations have a large number of parameters, and each patient, real or virtual, has a unique set of values for these parameters, referred to as their \textit{patient profile}.

The information flow of the APS simulation follows that of the real APS (see Fig. \ref{fig:controller} and Fig. \ref{fig:closeloop}). The patient model receives insulin dosage from the controller through the insulin pump, and the controller receives BG data through the CGM. Different control algorithms can be implemented without affecting the flow of information to and from the patient model. 

Improved control algorithms have driven recent advances in APS performance. Traditional control algorithms such as proportional integral derivative (PID) \cite{hu_improved_2015}, model predictive control (MPC) \cite{turksoy_integrated_2014}, and fuzzy logic \cite{nimri_md-logic_2017} have given way to machine learning based techniques such as deep neural networks \cite{sun2018predicting,govindaraju2022event} and reinforcement learning \cite{emerson_offline_2022, shaban-nejad_personalized_2021}, which offer more powerful insights into relationships in sensor data. However, machine learning techniques are far more intractable than traditional algorithms, so they must be rigorously tested before deployment to ensure the ML controller performs accurately in any possible scenario. This gives rise to a particular need for comprehensive testbeds in APS.

\textbf{Safety of APS:} 
A shortcoming of current testbeds is a lack of inclusion of adverse events that happen rarely and outside of standard system behavior. Safety-critical medical devices such as APS are often worn by or implanted in the patients, requiring them to operate on wireless networks and within unpredictable environments and activity settings, which naturally leads to safety concerns. For example, a smartphone app was recently developed for APS to remotely interface with CGM and insulin pump devices \cite{deshpande_design_2019}.
Several past studies have shown that medical devices (e.g., patient monitors, infusion pumps, implantable pacemakers, and tele-operated surgical robots) are vulnerable to accidental faults or malicious attacks with potential adverse impacts on patients \cite{alemzadeh_resiliency_2012, alemzadeh2013analysis,alemzadeh2016adverse,xu2019analysis,zhou2021data,zhou2022robustness}.
% Faults, or adverse events, are disruptions in standard APS function that put the health of the patient at risk. 

In APS, accidental faults or malicious attacks might happen in any of the system components \cite{ramkissoon2017review}, including the CGM \cite{Facchinetti2014ModelingTG}, insulin pump \cite{li2011hijacking}, or the APS controller \cite{khan2020runtime}. %and result in adverse events (involving device malfunctions, patient injuries, or deaths). %hypoglycemic events, hyperglycemic events
Since a variety of CGMs and insulin pumps are currently available in commercial diabetes management systems \cite{lal_realizing_2019}, large amounts of real-world data on faults and security vulnerabilities that led to recalls and adverse events are available from the public databases for analysis and simulation. 
We searched the FDA recalls~\cite{recall2022} and manufacturer and user facility device experience (MAUDE) databases~\cite{MAUDE2022} for safety issues related to APS during the last ten years. As shown in Table \ref{tab:recall}, over this period, millions of APS devices were recalled globally, and millions of adverse events (involving device malfunctions, patient injuries, or deaths) were reported by diabetic patients, healthcare professionals, and device manufacturers. This indicates an urgent need to investigate and improve the safety and dependability of APS devices. % (e.g., insulin pumps, APS control algorithms, and CGMs). 
Note that a single recall event corresponding to a software or hardware defect might lead to the removal or upgrade/repair of all the devices on the market with that software or hardware component. Also, the root causes of adverse events cannot be concluded solely based on the number of reports and the limited information available in the public FDA databases~\cite{MAUDE2022}. %and should be further investigated by the manufactcan be only determined after 
%~before the investigation by the company.
%
%MDR data alone cannot be used to establish rates of events, evaluate a change in event rates over time or compare event rates between devices. The number of reports cannot be interpreted or used in isolation to reach conclusions about the existence, severity, or frequency of problems associated with devices. Confirming whether a device actually caused a specific event can be difficult based solely on information provided in a given report. Establishing a cause-and-effect relationship is especially difficult if circumstances surrounding the event have not been verified or if the device in question has not been directly evaluated. MAUDE data does not represent all known safety information for a reported medical device and should be interpreted in the context of other available information when making device-related or treatment decisions.

Previous clinical trial studies have also shown the occurrence of adverse events due to pump infusion set failures, characterized by patterns of increasing glucose values despite increased insulin infusion \cite{cescon2016early}. Examples of adverse events with the risk of harm to patients are severe hypoglycemia, diabetic ketoacidosis (serum glucose > 250 $mg/dL$ \cite{westerberg2013diabetic}), serious events related to the device, hyperglycemia or ketosis without diabetic ketoacidosis ~\cite{brown2019six}.

% \usepackage{graphicx}
\begin{table}[t]
\caption{Recalls and Adverse Events of APS Devices (2012$-$2021)}
\vspace{-1em}
\label{tab:recall}
\resizebox{\columnwidth}{!}{%
\begin{tabular}{|l|l|l|l|}
\hline
\textbf{APS Component}      & \textbf{No. Recalls} & \textbf{No. Products}& \textbf{No. Adverse Events} \\ [0.5ex] \hline 
\textbf{Glucose Monitors} & 48                                    & 5.25 million & 1.62 million          \\ [0.5ex]\hline
\textbf{Insulin Pumps}    & 44                                   & 2.45 million  &1.11 million            \\ [0.5ex] \hline
\end{tabular}%

% \begin{tabular}{|l|l|l|l|l|}
% \hline
% \textbf{APS Device}      & \textbf{\#Recall} & \textbf{Recall Class} & \textbf{\#Products}& \textbf{\#Adverse Event} \\ \hline
% \textbf{Glucose Monitor} & 62                  & 1,2                   & 115.12 million & 1.62 million          \\ \hline
% \textbf{Insulin Pump}    & 44                  & 1,2,3                 & 2.99 million  &1.11 million            \\ \hline
% \end{tabular}%
}
\vspace{-2em}
\end{table}

\section{Design of Closed-loop APS Testbed}

\begin{figure}[t]
    \centering
    \includegraphics[width=0.85\columnwidth]{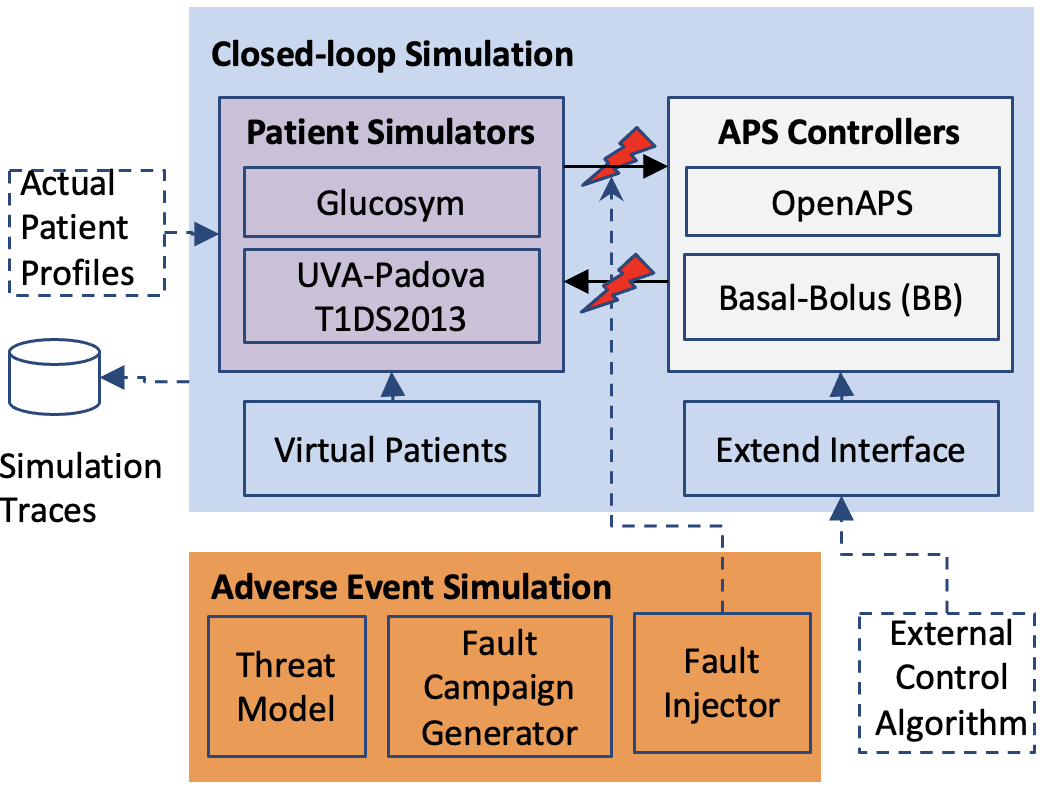}
    \vspace{-1em}
    \caption{Overall Structure of the Closed-loop APS Testbed}
    \vspace{-1.5em}
    \label{fig:closeloop}
\end{figure}

The overall structure of the open-source closed-loop Artificial Pancreas System (APS) testbed is shown in Fig. \ref{fig:closeloop}. The APS testbed includes two state-of-the-art glucose simulators (Glucosym simulator \cite{Glucosym} and the UVA-Padova Type 1 Diabetes simulator \cite{man_uvapadova_2014}) and two control software (OpenAPS and Basal-Bolus), together with 40 virtual patients. 
The simulator can run with the integrated virtual patient library or by loading actual patient profiles. Similarly, the testbed also includes an extending interface to the controllers that can load external control algorithms to help improve or evaluate the controllers in commercial insulin pumps. Note that only one simulator and controller are selected to run the closed-loop simulation. We also design an adverse event simulator that can emulate common adverse events in APS, including hypoglycemic events, hyperglycemic events, diabetic ketoacidosis, or other device malfunctions (e.g., in CGM sensors, insulin pumps, or controllers), by injecting faults into the input/output of the control software at compile time. 

The proposed closed-loop APS testbed and generated data traces are made publicly available to the research community\footnote{\url{https://github.com/UVA-DSA/APS_TestBed}}. The testbed is implemented with Python programming language at the application level, and can be installed on a Ubuntu operating system (16.04 LTS at least) automatically with an auto-script. This testbed offers a platform for other researchers to evaluate the performance of different control algorithms, validate the efficiency or safety of insulin delivery, develop the safety assurance or monitoring mechanisms for APS, and investigate the application of machine learning techniques in Type 1 diabetes treatment. The following subsections present a detailed description of the different components in the testbed. 
%This closed-loop simulation platform and generated data traces are made publicly available to the research community.

% The particular APS system we considered here is OpenAPS~\cite{lewis2020yourself}. To evaluate the effect of the system on patients through simulation, we developed a closed-loop system by integrating the OpenAPS control software with the Glucosym patient simulator \cite{Glucosym} (Fig. \ref{fig:closeloop}). 

\subsection{Patient Glucose Simulators}

% We integrate two state-of-the-art glucose simulator into the APS testbed: Glucosym \cite{Glucosym} and UVA-Padova Type 1 Diabetes simulator \cite{man_uvapadova_2014}.  
Table \ref{tab:summary-simulator} shows an overview of the dynamic models used by each glucose simulator to emulate the effect of insulin dosage on the body, along with the required parameters for characterizing the patient profiles to run the simulators. 

\begin{table}[t]
\caption{Summary of Patient Glucose Simulators}
\vspace{-1.5em}
\label{tab:summary-simulator}
\resizebox{\columnwidth}{!}
{
\centering
\begin{threeparttable}

\begin{tabular}{|c|l|c|}
\hline
\multicolumn{1}{|l|}{\textbf{Simulator}}                                        & \textbf{Dynamic Model}                 & \multicolumn{1}{l|}{\textbf{\begin{tabular}[c]{@{}l@{}}Patient\\ Profiles\end{tabular}}} \\ \hline
\multirow{6}{*}{\textbf{Glucosym}}                                              & Medtronic Virtual Patient (MVP) Model: & \multirow{6}{*}{\begin{tabular}[c]{@{}c@{}} C$_{I}$, $\tau$$_{1}$, \\
$\tau$$_{2}$, V$_{G}$,\\
p$_{2}$, $EGP$, \\
$GEZI$, S$_{I}$ \end{tabular}}                                                                        \\
                                                                                & sub-cutaneous insulin delivery,        &                                                                                          \\
                                                                                & the plasma insulin concentration,      &                                                                                          \\
                                                                                & the insulin effect,                    &                                                                                          \\
                                                                                & the glucose kinetics,                  &                                                                                          \\
                                                                                & and the glucose appearance.            &                                                                                          \\ \hline
\multirow{4}{*}{\textbf{\begin{tabular}[c]{@{}c@{}}UVA-\\ Padova\end{tabular}}} & Model of Kovatchev et al. \cite{Kovatchev2009insilico}:    & \multirow{4}{*}{\begin{tabular}[c]{@{}c@{}} $EGP$,\\ $U_{ii}$, $U_{id}$, \\$k_1$, $k_2$, \\$G_{pb}$ \end{tabular}}                                                                        \\
                                                                                & plasma concentration,                  &                                                                                          \\
                                                                                & glucose fluxes,                        &                                                                                          \\
                                                                                & and insulin fluxes.                    &                                                                                          \\ \hline
\end{tabular}%

% \begin{tabular}{|>{\centering\arraybackslash}m{1.35cm}|>{\raggedright\arraybackslash}m{5.1cm}|>{\centering\arraybackslash}m{1.25cm}|}
% \hline
% \textbf{Simulator} & \textbf{Dynamic Model} & \textbf{Patient Profiles} \\ \hline

%  Glucosym & \makecell{\pbox{5.1cm}{Medtronic Virtual Patient (MVP) Model:\\
% sub-cutaneous insulin delivery,\\ the plasma insulin concentration,\\ the insulin effect,\\ the glucose kinetics,\\ and the glucose appearance.}}
% & C$_{I}$, $\tau$$_{1}$, $\tau$$_{2}$, V$_{G}$, p$_{2}$, $EGP$, $GEZI$, S$_{I}$        \\ \hline

% % \\ Glucose minimal model \cite{bergman1979quantitative}
% UVA-Padova  & \makecell{\pbox{5.1cm}{Model of Kovatchev et al. \cite{Kovatchev2009insilico}:\\ plasma concentration,\\ glucose fluxes, \\and insulin fluxes.}}         &    $EGP$, $U_{ii}$, $U_{id}$, $k_1$, $k_2$, $G_{pb}$  \\ \hline
% \end{tabular}

\begin{tablenotes}
% \normalfont
%  \item[*] BW=Body Weight (kg).
 \item[*]C$_I$=Insulin clearance (dL/min).
 \item[*]{$\tau$$_1$, $\tau$$_2$=Time constant associated with insulin movement between the SC delivery site and plasma (min).}
 \item[*]V$_G$=Distribution volume in which glucose equilibrates (dL).
 \item[*]p$_2$=Delay in insulin action upon increase in plasma insulin (1/min).
 \item[*]  EGP=Endogenous glucose production rate that would be estimated at zero insulin (mg/dL/min).
 \item[*] GEZI=Effect of glucose per se to increase glucose uptake into cells and lower endogenous glucose production at zero insulin (1/min).
 \item[*] S$_{I}$=Baseline sensitivity factor (dl/micro Unit).
 \item[*] $U_{ii}$=Insulin-independent glucose utilization.
 \item[*] $U_{id}$=Insulin-dependent glucose utilization.
 \item[*] $k_1$, $k_2$=Rate parameters of glucose kinetics.
 \item[*] $G_{pb}$=Initial amount of glucose in plasma.
\end{tablenotes}

\end{threeparttable}
}
\vspace{-2em}
\end{table}

\textbf{Glucosym Patient Simulator:}
\label{sec:Glucosym}
The Glucosym simulator is an open-source human body glucose simulator that was developed to help build and test automatic insulin delivery systems. 
This simulator contains patient models derived from data collected from 10 actual adult patients with Type I diabetes mellitus for 18 $\pm$ 13.5 years aged 42.5 $\pm$ 11.5 years, with their glucose dynamics predicted using a Medtronic virtual patient (MVP) model \cite{KanderianT1}.

The MVP model includes five components that describe the sub-cutaneous insulin ($I_{SC}$) delivery, the plasma insulin concentration ($I_{P}$), the insulin effect ($I_{EFF}$) to lower blood glucose, the glucose kinetics, and the glucose appearance following a meal ($R_A$) (see Eq. \ref{eq:eq1}-\ref{eq:eq5}).
A three-compartment model \cite{Insel1974modeling} was used to identify the insulin activity after injection to the patient body (see Eq. \ref{eq:eq1}-\ref{eq:eq3}). 
With the value of glucose appearance given by the two-compartment model shown in Eq. \ref{eq:eq5}, the Bergman minimal model \cite{Bergman-Model} and Sherwin model \cite{Sherwin-Model} described in Eq.\ref{eq:eq4} were finally used to derive an estimation of the BG value at the next step. 
These five equations form the basis of the MVP dynamic model used in the Glucosym simulator for educating and training individuals with Type 1 diabetes \cite{KanderianT1}:

\begin{align}
\frac{dI_{SC}(t)}{dt}&=-\frac{1}{\tau_{1}}\cdot\left( I_{SC}(t)-\frac{ID(t)}{C_{I}}\right)  \label{eq:eq1}\\
\frac{dI_{P}(t)}{dt}&=-\frac{1}{\tau_{2}}\cdot(I_{P}(t)-I_{SC}(t))\label{eq:eq2}\\
\frac{dI_{EFF}(t)}{dt}&=-p_2\cdot(I_{EFF}(t)-S_I\cdot I_{P}(t))\label{eq:eq3}\\
\frac{dBG(t)}{dt}&=-(GEZI+I_{EFF}(t))\cdot BG(t)+EGP+R_{A}(t)\label{eq:eq4}\\
R_{A}(t)&=\frac{C_{H}(t)}{V_{G}\cdot{\tau_{m}}^2}\cdot t\cdot e^{-\frac{1}{\tau_{m}}} \label{eq:eq5}
\end{align}

where, $GEZI, EGP, S_I, C_I, p2, \tau1, \tau2$ are patient-specific parameters, with their explanation presented in Table \ref{tab:summary-simulator}. Other parameters, such as the input information of insulin doses and sampling frequency, are also needed for running the Glucosym simulator. The full list of input parameters used in this simulator is listed in Table \ref{tab:glucosym_Input}. An implementation of this simulator is publicly available at \cite{Glucosym}.

\begin{table}[t]
% \vspace{-1em}
\begin{center}
%\vspace{-1em}
\caption{Input Parameters of Glucosym Simulator.}
\vspace{-1em}
\label{tab:glucosym_Input}
\resizebox{\linewidth}{!}{
\begin{tabular}{ | m{0.8cm} | m{7.2cm} |} 
\hline
\makecell{\textbf{Input}} &\textbf{Description}  \\ 
\hline
Insulin Dose & Insulin dose in units given during the time-step. In the case of a basal (insulin delivery) adjustment, we need to calculate how much insulin will be given in the time-step defined by "dt" (i.e. how many insulin units will be given in 5 minutes by the set basal profile or temporary basal?).  \\ 
\hline
\makecell{dt} & Change in time each step in minutes. \\
\hline
\makecell{Index} & Current index from the start of the simulation, starting at 0.\\
\hline
\makecell{Time} & Total simulation run-time in minutes.\\
\hline
\makecell{Basal} & The delivery of insulin. \\
\hline
\makecell{Events} & Events are set so that the simulator will consider them during the run. The events were sent on-the-go.\\
\hline
\end{tabular}}
\end{center}
\vspace{-1.5em}
\end{table}

\begin{figure*}[]
    \centering
     \begin{minipage}{0.9\columnwidth}
	\scriptsize \centering
    \includegraphics[width=\columnwidth]{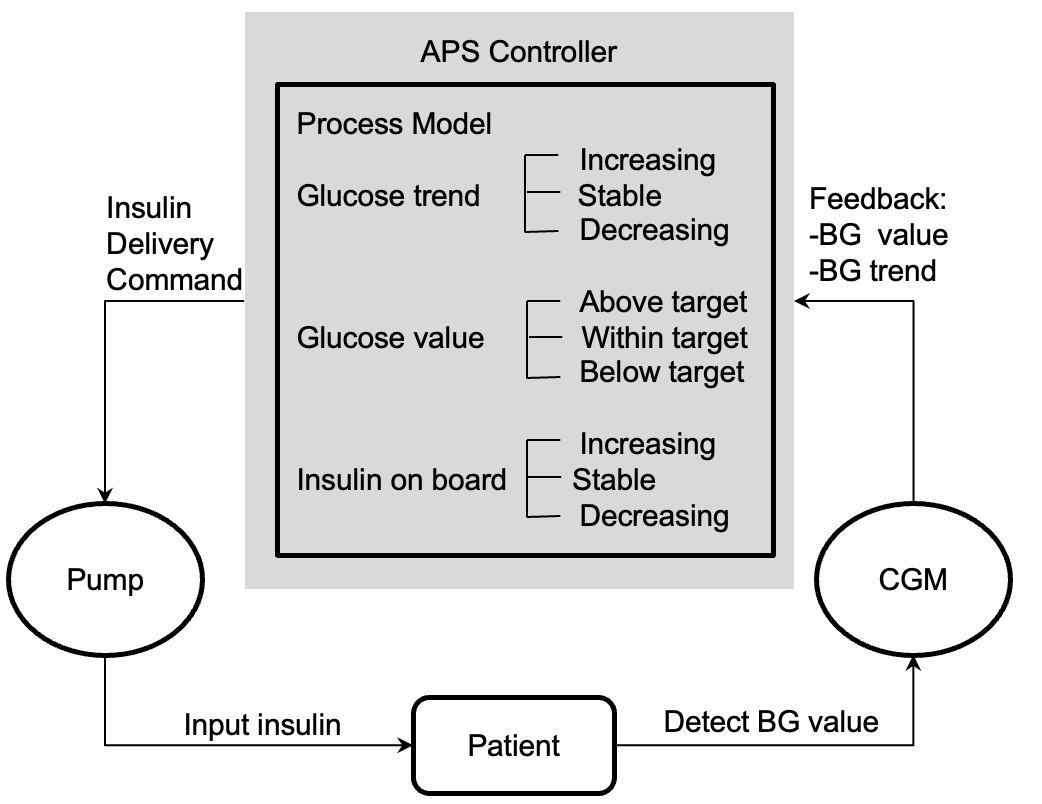}
    
    % (a)  
    \end{minipage}
    %\hfill
    \begin{minipage}{0.9\columnwidth}
	\scriptsize \centering
    \includegraphics[width=\columnwidth]{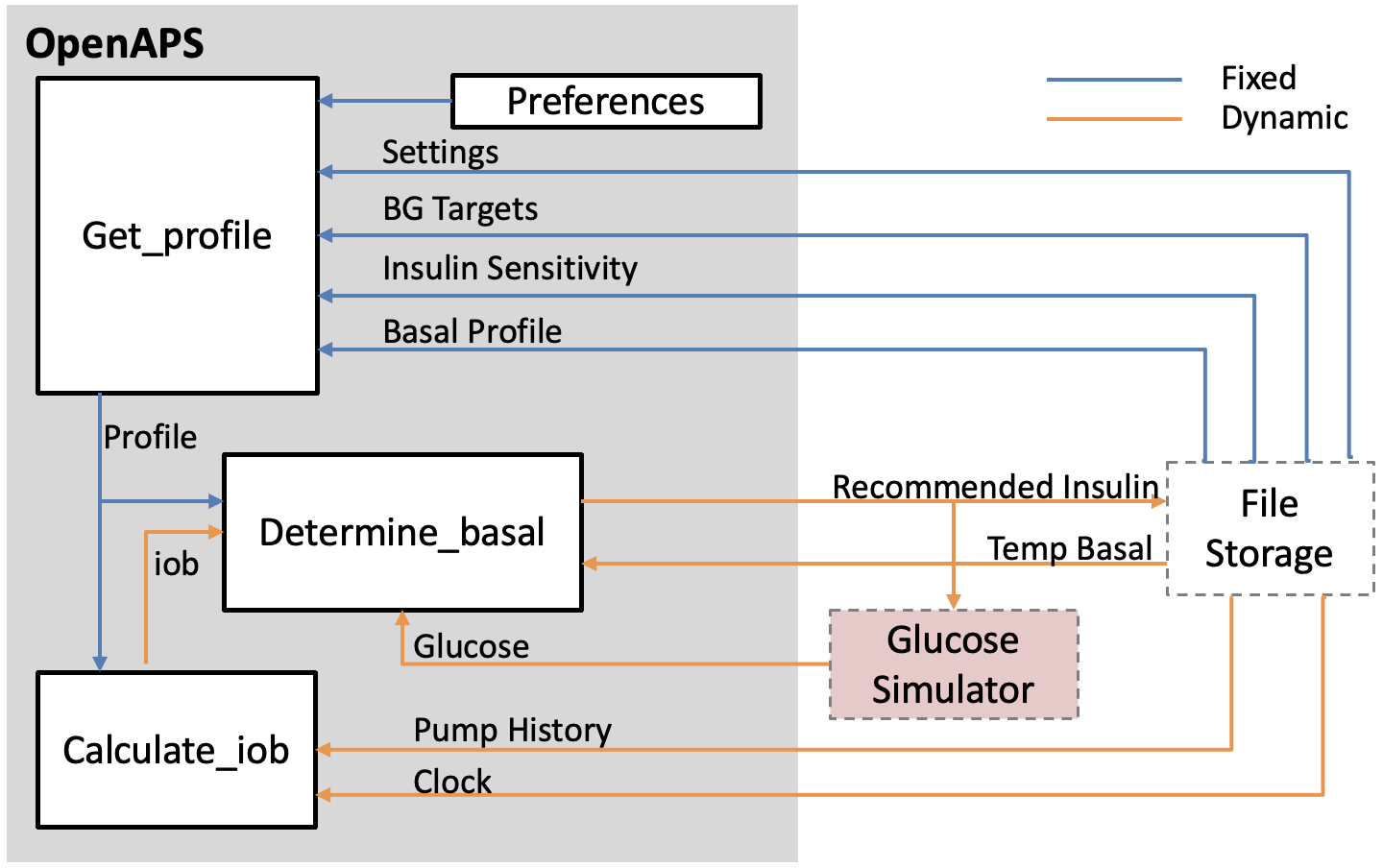}
    
    % (b)  
    \end{minipage}

    \vspace{-1em}
    \caption{Typical APS Control Structure (Left) and OpenAPS Architecture and Input/Output (Right).}
    %\todo[inline]{Do we still need this figure?}
    %\todo[inline, color=green!40]{We can keep this}
    \label{fig:OpenAPS_arch}
    \vspace{-1.5em}
\end{figure*}

\textbf{UVA-Padova simulator:} The other simulator we integrated into the APS testbed is the UVA-Padova Type 1 Diabetes Simulator, which FDA has approved for pre-clinical testing on animals. 
In this simulator, the model of glucose kinetics is described using the following equations \cite{man_uvapadova_2014}:

\begin{align}
    \frac{dG_{p}(t)}{d_t} &= EGP - U_{ii} - k_{1}G_{p}(t)+k_{2}G_{t}(t) \; , \; G_{p}(0)=G_{pb}\\
     \frac{dG_{t}(t)}{d_t} &= U_{id}(t) + k_{1}G_{p}(t)-k_{2}G_{t}(t) \; , \; G_{t}(0)=G_{pb}\frac{k1}{k2}
\end{align}

\noindent where $G_{p}(t)$ represents the amount of glucose in plasma, and $G_{p}(t)$ describes the amount of glucose in the tissue. The blood glucose level that the CGM samples is given by Equation \ref{uva_sim_bg}:

\begin{equation} \label{uva_sim_bg}
    G(t) = \frac{G_p(t)}{V_g}
\end{equation}

The endogenous glucose production rate, $EGP$, is modeled as a function of glucose in plasma, $G_p(t)$, and delayed insulin action in the liver, $X^L(t)$, as shown in Equation \ref{eqn:uva_sim_egp}.

\begin{equation} \label{eqn:uva_sim_egp}
    EGP = k_{p1} - k_{p2}\cdot G_p(t) - k_{p3}\cdot X^L(t)
\end{equation}

$X^L(t)$ is based on insulin concentration in plasma. The insulin dose delivered to the patient by the pump, $ID(t)$, factors into this plasma insulin level via the insulin subsystem, which is split into $I_k(t)$ and $I_{sc}(t)$. $I_{sc}(t)$ represents the subcutaneous insulin level, and is impacted by insulin doses as follows:

\begin{table}[b]
% \vspace{-2em}
\begin{center}
\caption{\label{tab:T1DS_Input} Input Parameters of UVA-Padova Simulator.}
\vspace{-1em}
\resizebox{\linewidth}{!}{
\begin{tabular}{ | p{2cm} | p{6cm} |} 
\hline
\textbf{Input} & \textbf{Description}  \\ 
\hline
Initial BG & Starting value for patient's blood glucose\\
\hline
Sensor Settings & Type of CGM sensor and associated settings\\
\hline
Pump Settings & Type of insulin pump and associated settings\\
\hline
Meals & Sequence containing the time and size of each meal during the simulation\\
\hline
Profile & Unique parameters for the patient profile\\
\hline
Start Time & Beginning time for the simulation\\
\hline
Seed & Random number generator seed used for noise in sensor readings, etc.\\
\hline
Insulin Dose & Insulin dose to give to the patient for each step\\
\hline
\end{tabular}
}
\end{center}
% \vspace{-2em}
\end{table}

\begin{equation}
    \frac{dX^L(t)}{dt} = -k_i \cdot \left( X^L(t) - k_{ai}I_k(t) - k_{bi}I_{sc}(t)  \right)
\end{equation}

\begin{equation}
    \frac{dI_{sc}(t)}{dt} = k_{sc}I_{sc}(t) + ID(t)
\end{equation}

Other variables in the above equations are constant rate parameters that are part of the patient profile. This model was improved in 2013 by implementing the notion that insulin-dependent utilization increases non-linearly when glucose decreases below a certain threshold. Similar to the Glucosym simulator, the UVA-Padova simulator also uses the minimal glucose model to couple insulin action on glucose utilization and production. 
Other parameters required by the UVA-Padova simulator to run regularly are listed in Table \ref{tab:T1DS_Input}.

% The first UVA-Padova simulator was approved by the FDA in 2008 [3] for a singlemeal scenario only. 
% The VPs were represented by a set of model parameters which were extracted randomly from joint distributions of parameters. 
% A new version was published in 2014 [52], in which improved glucose kinetics in hypoglycemia and glucagon dynamics were implemented. The virtual population was also improved in terms of clinical parameters such as carbohydrates ratio and correction factor. This version was also approved

The two glucose simulators integrated with the APS testbed could also handle a single meal scenario for the virtual patient (VP) population, which is challenging for regulating BG in Type 1 diabetes because of unexpected human activities (e.g., meals or exercises) and patient variability (inter-patient and intra-patient).

% Using this simulator, one can compare multiple algorithms with the same patient model, test on a patient profile with parameter variations during the day, or test against a population of patients. Sample parameters of this simulator are shown in Fig. \ref{fig:Glucosym_Param}.

\subsection{APS Controllers}
We integrate two typical control algorithms into the APS testbed: a PID-based OpenAPS controller and a Basal-Bolus controller.

% We integrated two State-of-the-art APS control algorithms: OpenAPS and Basal-Bolus in the propose test tool. 

% \begin{comment}
\begin{table}[b!]
\begin{center}
\vspace{-1.5em}
\caption{\label{tab:OpenAPS_Input} Input Parameters of OpenAPS.}
\vspace{-1em}
\resizebox{\linewidth}{!}{
\begin{tabular}{ | p{2cm} | p{6cm} |} 
\hline
\textbf{Input} & \textbf{Description}  \\ 
\hline
Settings & Various settings specific to the pump  \\ 
\hline
BG targets & High/low glucose targets set up in the pump  \\ 
\hline
Insulin Sensitivity & The expected decrease in BG as a result of one unit of insulin \\ 
\hline
Basal profile & The basal rates that are set up in the pump \\ 
\hline
Preferences & User-defined preferences \\ 
\hline
Pump history & Last 5 hours data directly from the pump \\ 
\hline
Clock & Date and time that is set on the pump \\ 
\hline
Temp\_basal & Current insulin delivery rate set up in pump \\ 
\hline
Glucose & Glucose level sensed by CGM \\ 
\hline
\end{tabular}
}
\end{center}
\end{table}
% \end{comment}

\textbf{OpenAPS} is an advanced open-source control software used in the diabetes DIY community \cite{lewis2020yourself} that has comparable results with more rigorously developed and tested AP systems for glycaemic control \cite{melmer2019glycaemic} and is far safer than standard pump/CGM therapy with no reports of severe hypo- or hyperglycemic events \cite{lewis2016real}.

The OpenAPS adjusts the insulin delivery of an infusion pump to automatically keep the BG level of the diabetic patient within a safe range. The internal architecture and necessary input-output connections of OpenAPS are shown in Fig. \ref{fig:OpenAPS_arch}. The description of input parameters is listed in Table \ref{tab:OpenAPS_Input}. 
The shaded region indicates the OpenAPS controller, and the "File Storage" section reflects the behavior of the insulin pump. The functionality of OpenAPS can be divided into three processes. The \textit{Get\_profile} process accepts pump settings, target BG (BGT), insulin sensitivity, basal profile, and preferences as inputs and creates a profile required to calculate both IOB and recommended insulin delivery. The \textit{Calculate\_iob} process gets profile, clock, and pump history as input and calculates IOB.
Finally, the \textit{Determine\_basal} process accepts the profile, IOB, BG, and current insulin delivery (temp\_basal) and calculates the suggested insulin delivery to the patient.

\begin{algorithm}[b]
 \caption{OpenAPS Algorithms}
 \label{algcloseloop}
  \uIf{BG is rising, but $eventual BG<BG\_Target$}{
   cancel any temp basal\;
   }
   \uElseIf{BG is falling, but $eventualBG>BG\_Target$}{
   cancel any temp basal\;
  }
  \ElseIf{ $eventualBG>BG\_Target$}{
   cancel 30min temp basal\;
       \uIf{recommended temp>existing basal}{
       issue the new high temp basal\;
       }
       \uElseIf{recommended temp<existing basal}{
       issue the new high temp basal\;
      }
      \ElseIf{0 temp for >30m is required}{extend zero temp by 30 min\;
      }
  }
\end{algorithm}

More specifically, OpenAPS collects the previously delivered insulin amount, combined with the duration of the activity, and it calculates the net IOB. Using the glucose sensor readings, OpenAPS then calculates the eventual BG using the following equation \cite{determine-basal}:
    % \vspace{-0.5em}
\begin{equation}
    % \vspace{-0.5em}
    eventual BG = Current BG - ISF*IOB + deviation 
    % \vspace{-0.5em}
\end{equation}
    \vspace{-1em}

where \(Current BG\) is the current BG,  \(ISF\) is the Insulin Sensitivity Factor, % \(IOB\) is the Insulin on Board 
and \(EventualBG\) is the estimated BG by the end of current insulin delivery. 
A \textit{deviation} term is also added, which is the difference in BG prediction based on purely insulin activity.
%Fig. \ref{fig:openapsBGpred} shows examples of how OpenAPS predicts the BG traces in different scenarios.

While the current BG is below a threshold value, OpenAPS continues to issue a temporary zero insulin delivery until the BG rises. Otherwise, OpenAPS determines whether the glucose values rise or fall more than expected. In that case, it performs the course of actions shown in Algorithm \ref{algcloseloop} \cite{determine-basal}.

% \begin{figure}[ht]
%     \centering
%     \includegraphics[width=8cm, height=5cm]{figures/openAPS_Algorithm.png}
%     \caption{OpenAPS Algorithm}
%     \label{fig:algorithm1}
% \todo[inline]{line 2 baasal -> basal}
%     \vspace{-1em}
% \end{figure}

% \begin{comment}
\begin{table}[t!]
\begin{center}
\caption{\label{tab:BB_Input} Input Parameters of Basal-Bolus Controller.}
\vspace{-1em}
\begin{tabular}{ | p{1cm} | p{6cm} |} 
\hline
\textbf{Input} & \textbf{Description}  \\ 
\hline
CGM & Continuous glucose monitor sensor reading  \\ 
\hline
CHO & Grams of carbohydrates consumed by patient (if meal occurred at current step)  \\ 
\hline
BW & Patient's body weight \\
\hline
$u_{2ss}$ & Steady state insulin rate per kilogram \\
\hline
CR & Insulin to carbs ratio \\
\hline
CF (ISF) & {Insulin correlation (sensitivity) factor\cite{CF&ISF}}\\
\hline
\end{tabular}
\end{center}
\vspace{-1.5em}
\end{table}
% \end{comment}

\textbf{Basal-Bolus} regimens are widely used in insulin pumps \cite{BBcontroller, brown2019six, cameron_closed-loop_2011}. 
Basal provides a constant supply of insulin to bring down high resting blood glucose levels.
Bolus insulin, on the other hand, has a much more powerful but shorter-lived effect on blood sugar, making it an ideal supplement for people with diabetes to take after meals and in moments of extremely high blood sugar.

% \usepackage{graphicx}
%\begin{tabular}[c]{@{}l@{}}
\begin{table*}[b!]
\caption{Example Recall Event Reports That Involved Device and Software Malfunctions.}
\vspace{-1em}
\label{tab:recall-list}
\resizebox{\textwidth}{!}
{%
\begin{tabular}{|p{1.75cm}|p{15cm}|p{2.5cm}|p{1.5cm}|}
\hline
\textbf{Recall ID} & \textbf{Summary Recall Description}                                                                                                                                                   & \textbf{FDA Determined Cause} & \textbf{Affected Device} \\ \hline
Z-1074-2013 & The blood glucose meter will shut off and revert to set up mode at glucose values above 1023 instead of displaying EXTREME HIGH GLUCOSE. & Software Design               & Glucose Monitor                                                         \\ \hline
Z-1034-2015 & Calibration factors in the pump are overwritten during a programming step. The force sensor could send a lower signal value to the pump processor. & Software Design & Insulin Pump\\ \hline
Z-1734-2015 &
    If the user does not act upon the E6 and E10 error messages appropriately, insulin delivery will be stopped and, if unnoticed, may lead to severe hyperglycemia.
    & Device Design & Insulin Pump \\ \hline
Z-1359-2012 &
    An error was discovered in the blood glucose meter software so that the meter turns itself off when a user attempts to view results in the "Results Log" when the log has 256 or a multiple of 256 items to display.
    & Software Design & Glucose Monitor \\ \hline
Z-0929-2020 &
     The mobile receiver can become stuck on the initialization screen when powering on. This will cause patients not to be able to receive glucose values or alerts
    & Software Design & Glucose Monitor\\ \hline
%Z-1376-2012 &
    %There is an error in the pump software where the insulin pump did not allow the user to save changes to the time and date on the pump if those changes were made on a leap day.
    %& Software Design & Insulin Pump \\ \hline
Z-1562-2020 &
    The company identified potential interference from hydroxyurea. Patient use of the anti-neoplastic drug may falsely elevate glucose readings on the CGM.
    & Under Investigation & Gluocse Monitor\\ \hline
Z-2165-2020 &
    After the device has been in use for about two months, data processing in the PDM can be slowed such that the Bolus Calculator fails to accurately subtract the correct amount of IOB before suggesting a bolus amount.
    & Device Design & Insulin Pump \\ \hline
Z-1772-2021 &
    Under certain conditions, a software fault is detected when a large bolus delivery at a quick bolus speed completes. If the user is unaware of the amount of active insulin and delivers an additional bolus, there is a risk of insulin over delivery.
    & Software Design & Insulin Pump \\ \hline
\end{tabular}%
}
\end{table*}

In the Basal-Bolus (BB) Controller, the constant supply of basal insulin is determined as shown in Equation \ref{bb_basal} \cite{viroonluecha2021evaluation}:
\vspace{-0.5em}
\begin{equation} 
\label{bb_basal}
    I_{basal} = \frac{u_{2ss}\cdot BW}{6000}
\end{equation}

\noindent where $u_{2ss}$ is the patient's steady-state insulin rate per kg and $BW$ is body weight (kg), meaning basal insulin is in units of insulin per minute. Bolus insulin is determined by Equation \ref{bb_bolus} when a meal has occurred (otherwise, no bolus is given) \cite{viroonluecha2021evaluation}:
% {
% \begin{equation} \label{bb_bolus}
%     I_{bolus} = \frac{CHO}{CR} + \left(BG > 150\right) \frac{BG - BG_{target}}{CF}
% \end{equation}
% }

\begin{equation} \label{bb_bolus}
    I_{bolus} =
    \begin{cases}
    \frac{\displaystyle CHO}{\displaystyle CR} & \text{if } BG \leq 150
    \\[5pt]
    \frac{\displaystyle CHO}{\displaystyle CR} + \frac{\displaystyle BG - BGT}{\displaystyle CF} & \text{if } BG > 150\\
    \end{cases}
\end{equation}
\noindent where $CHO$ is the meal's size in grams of carbohydrates, $BG$ is the CGM sensor reading, $BGT$ is the target blood glucose of 120, $CR$ is the insulin to carbs ratio, and $CF$ is the correlation factor. The list of input parameters of the Basal-Bolus controller is also summarized in Table \ref{tab:BB_Input}. This bolus is the units of insulin to be delivered, so it is divided by the length of a simulation step to become units of insulin per minute.

% The other APS controller used is basal-bolus controller. This controller does not predict future BG value, but merely deliver the underlying basal rate set in the pump.

\subsection{Closed Loop Simulation}
% how it works together
Fig. \ref{fig:OpenAPS_arch} shows an example of the closed-loop simulation process by integrating the Glucosym simulator and OpenAPS control software. 
At each control loop, the estimated glucose value is updated and reported to the APS controller, based on which the controller calculates the recommended insulin dosage and sends it to the glucose simulator. The insulin amount is divided by 60 to convert the units from $Unit/hour$ to $Unit/minute$ to make OpenAPS and Glucosym work appropriately in a closed loop. The glucose value updates every five minutes (this is the value normally set by CGM \cite{fico2017exploring}), and so does the control action. 
% The recommended insulin dosage by the APS controller is sent to the simulator and the estimated glucose value is sent back to the controller. 
% write a section on how the testbed works, how the CGM and pump are simulated, with what frequency the APS receives glucose and sends insulin commands. How this resembles a real world case.

In the UVA-Padova simulation, the CGM sensor is simulated by looking up the subcutaneous glucose state variable in the patient model, applying noise, and clipping it to be within the range of values an actual CGM sensor can return. Similarly, the simulated pump receives a basal and a bolus input from the controller, converts the values into the appropriate units ($pmol/min$), and clips the inputs to be within the real range of the insulin pump before sending the values to the patient model. These calls occur once per minute (5 times per environment step).

% Does the control algorithm need patient-specific parameters other than target value?

The Basal-Bolus controller uses additional patient-specific parameters to calculate insulin doses. For the basal insulin, it requires the patient's body weight and steady-state insulin rate. For the bolus dose, it uses the patient's insulin to carbs ratio (CR) and correlation factor (CF).
% The BB controller also needed two extra parameters that were not in the patient profile: the insulin to carbs ratio (CR) and insulin correlation factor (CF).
Both CR and CF can be calculated from the Total Daily Dose (TDD) of insulin needed, which in turn is calculated from body weight, as shown in the following equations \cite{king_prospective_2007,man_uvapadova_2014}:
\vspace{-0.5em}
\begin{align}
    TDD &= 0.55 \cdot BW \label{bb_tdd} \\
    CR &= 450 / TDD \label{bb_cr} \\
    CF &= 1700 / TDD \label{bb_cf}
\end{align}

% how to calculate CR and CF 
% CR and CF can both be calculated from the patient's total daily dose of insulin (TDD), which in turn is based on the patient's body weight. TDD is 0.55 times body weight (in kg), CR is found by dividing 450 by the TDD , and CF is found by dividing 1700 by the TDD. \cite{man_uvapadova_2014}

\subsection{Adverse Event Simulator}
\label{sec:adversesimulator}
% Definition of adverse event in APS...
%An adverse event (AE) is any untoward medical occurrence in a patient or clinical investigation subject administered a pharmaceutical product, irrespective of the relationship between the adverse event and the device(s) or treatment \cite{AE}. This could happen due to malfunction of medical devices, accidental faults, or malicious attacks. Examples of adverse events involving APS include: malfunction~\cite{maude_malfunction}, injury~\cite{maude_injury}, and death~\cite{maude_death}.

After a medical device, such as a CGM, insulin pump, or APS, is distributed in the market, the FDA monitors reports of adverse events and other problems with the device and, when necessary, alerts health professionals and the public to ensure proper use of the device and safety of patients \cite{AE-FDA, alemzadeh2013analysis}.
A recall is a voluntary action that a device manufacturer takes to correct or remove from the market any medical devices that violate the laws administrated by the FDA \cite{RecallsDef}. Recalls are initiated to protect public health and well-being from devices that are defective or that present health risks such as disease, injury, or death. In rare cases, if the company fails to recall a device that presents a health risk voluntarily, the FDA might issue a recall order to the manufacturer. 

%The MAUDE database is a collection of adverse events of medical devices that volunteers, user facilities, manufacturers, and distributors reported to the FDA.
FDA regulations also require manufacturers to notify the FDA of the adverse events, including device malfunctions \cite{maude_malfunction}, serious injuries~\cite{maude_injury}, and deaths~\cite{maude_death} associated with medical devices. Not all reported adverse events lead to recalls. The device manufacturers and the FDA regularly monitor the adverse event reports to detect and correct problems in a timely manner.

% An adverse event (AE) can therefore be any unfavourable and unintended sign (including an abnormal laboratory finding), symptom, or disease temporally associated with the use of a medicinal (investigational) product, whether or not related to the medicinal (investigational) product

%We searched the database of FDA recall reports and the manufacturer and user facility device experience (MAUDE) and present the recall events and adverse events related to APS in Table \ref{tab:recall} during the recent 10 years. We see that millions of APS devices were recalled globally and millions of adverse events (including malfunction, injury, and death) have been reported by either the diabetic patients or manufacturers, indicating an urgent need to investigate/improve the safety and dependability of APS devices (e.g., insulin pumps, APS control algorithms, and CGMs). 

% Table \ref{tab:recall} shows the number of recall events and medical devices in the recent 10 years related to APS reported to FDA, such as stop of insulin delivery due to pump malfunction or adversary attack \cite{PumpRecall2-2019,PumpRecall2019,MedicalCyberattack2019}.

Table \ref{tab:recall-list} shows example recall events from the FDA database where malfunctions of the commercially available APS devices or software were reported.
The analysis and simulation of past recalls and typical adverse event scenarios can help with improving the design and test of the APS control algorithms and safety mechanisms and assessing their effectiveness in preventing similar adverse events~\cite{alemzadeh2013analysis, alemzadeh2015software, xu2019analysis}. However, it is too expensive and risky to simulate the adverse event scenarios with the actual patients and human operators in the loop due to the unacceptable consequences of adverse events and potential harm to patients. 
 
To better evaluate the resilience of APS control algorithms against such safety issues, we design an adverse events simulator integrated with the closed-loop simulation. Specifically, we design a software-implemented fault injection (SWFI) engine (see Fig. \ref{fig:closeloop}) that can automatically select a set of target locations within the APS software (e.g., variables representing the CGM sensor values and insulin dose commands) to inject faults (e.g., a zero value (\textit{Truncate}), a previous value (\textit{Hold}), or an arbitrary error value (\textit{Add/Sub})) and activate them under pre-defined trigger conditions and durations to mimic the typical adverse events listed in Table \ref{tab:threat_model}, including hyperglycemic (diabetic ketoacidosis) and hypoglycemic events, device malfunctions, and patient injuries. The adverse event simulator is an independent module and can be enabled or disabled manually. 

%DoS attack, availability attack, integrity attack, and hardware fault \cite{}.

\begin{comment}
Matching FDA recalls to simulated faults
Numbers correspond to rows in CGM/insulin csv files
Truncate:
    CGM: 7 8
    Pump: 26 30 31 32 33 46
    
    Total: 8
    
Hold:
    CGM: 2 7 73
    Pump: 29 31 32
    Both (CGM record): 3
    
    Total: 7
    
Add/Sub:
    CGM: 35 77 78 79
    Pump: 30 34 35 55 56 60 61 67 68 69 70
    
    Total: 15

certain condition caused inaccurate reading (specific): 12 23 24 25 29 30 41 90 91
incorrect performance (general): 13
wrong factory setting: 27 28 58 80 81 86 87 88 89
\end{comment}

% how the research community can use this closed-loop simulator.. Is it easy for them to download and use it?

\begin{table*}[tp!]
    %\vspace{-0.5em}
    \small
    \centering
    \caption{Simulated Fault and Adverse Event Scenarios}
    \vspace{-1em}
    % \resizebox{\textwidth}{!}
    {
    \begin{threeparttable}
    
    \begin{tabular}{|l|l|l|l|l|}
\hline
\textbf{Fault Type}                                                       & \textbf{Fault Injection Approach}                                                                                                                  & \textbf{Representative FDA Recalls}                                                                         & \textbf{No. Records} & \textbf{Possible Adverse Events}                                                                                                   \\ \hline
\multirow{2}{*}{Truncate}                                           & \multirow{2}{*}{Change output variables to zero value~\cite{li2011hijacking}\cite{ramkissoon2017review}}                                                                             & \multirow{2}{*}{\begin{tabular}[c]{@{}l@{}}Z-1074-2013, Z-1034-2015,\\ Z-1734-2015$\textsuperscript{1}$\end{tabular}} & \multirow{2}{*}{8}   & \multirow{6}{*}{\begin{tabular}[c]{@{}l@{}}Device Malfunction/\\ Hypoglycemia/ \\  Hyperglycemia/\\ Injury/\\ Death\end{tabular}} \\
                                                                    &                                                                                                                                    &                                                                                                  &                      &                                                                                                                                   \\ \cline{1-4}
\multirow{2}{*}{Hold}                                               & \multirow{2}{*}{Stop refreshing selected input/output variables~\cite{attack_pcs}\cite{ramkissoon2017review}}                                                                   & \multirow{2}{*}{\begin{tabular}[c]{@{}l@{}}Z-1359-2012, Z-0929-2020,\\ Z-1376-2012\end{tabular}} & \multirow{2}{*}{7}   &                                                                                                                                   \\
                                                                    &                                                                                                                                    &                                                                                                  &                      &                                                                                                                                   \\ \cline{1-4}
\multirow{2}{*}{\begin{tabular}[c]{@{}l@{}}Add/\\ Sub\end{tabular}} & \multirow{2}{*}{\begin{tabular}[c]{@{}l@{}}Add or subtract an arbitrary or particular value\\ to a targeted variable~\cite{attack_pcs}\cite{zhou2022Strategic}\end{tabular}} & \multirow{2}{*}{\begin{tabular}[c]{@{}l@{}}Z-1562-2020, Z-2165-2020,\\ Z-1772-2021\end{tabular}} & \multirow{2}{*}{15}  &                                                                                                                                   \\
                                                                    &                                                                                                                                    &                                                                                                  &                      &                                                                                                                                   \\ \hline
\end{tabular}%

    % \begin{tabular}{|p{1.5cm}|p{7cm}|p{3cm}|p{1.5cm}|p{3cm}|}
    % \hline
    % \textbf{Type}   & \textbf{Approach}& \textbf{Recall Examples}& \textbf{No. Records} & \textbf{Possible Adverse Event} \\ \hline
    
    % Truncate & Change output variables to zero value   \cite{li2011hijacking}\cite{ramkissoon2017review}& Z-1074-2013, Z-1034-2015, Z-1734-2015$\textsuperscript{1}$ & 8 &  Device Malfunction/     \\ \cline{1-4}
    
    % Hold            & Stop refreshing selected input/output variables\cite{attack_pcs}\cite{ramkissoon2017review}  & Z-1359-2012, Z-0929-2020, Z-1376-2012 & 7 &Hypoglycemia/ Injury/ \\
    % \cline{1-4}
    
    % % Max/min & Change the value of targeted variables to their maximum or minimum allowed values  \cite{attack_pcs} \cite{jha2019mlbased}& {Integrity attack \cite{li2011hijacking}\cite{FDApumprecall}/ } \\ %\cline{1-2}

    % Add/Sub         & Add or subtract an arbitrary or particular value to a targeted variable  \cite{attack_pcs}\cite{zhou2022Strategic}& Z-1562-2020, Z-2165-2020, Z-1772-2021 & 15 & Hyperglycemia/Death       
    % \\\hline
    % \end{tabular}
    
    \begin{tablenotes}
     \item[1] Recall IDs assigned by FDA which are also listed in Table \ref{tab:recall-list}.
    %  \item[1] Recall events number by FDA which can be searched for on \url{https://www.accessdata.fda.gov/scripts/cdrh/cfdocs/cfres/res.cfm}.
    \end{tablenotes}
    
    \end{threeparttable}

    }
    \label{tab:threat_model}
    \vspace{-1em}
\end{table*}

%\section{Validation of the Closed}
\section{Validity Assessment}

\begin{figure}[b!]
    \centering
    \vspace{-1.5em}
    \includegraphics[width=\columnwidth]{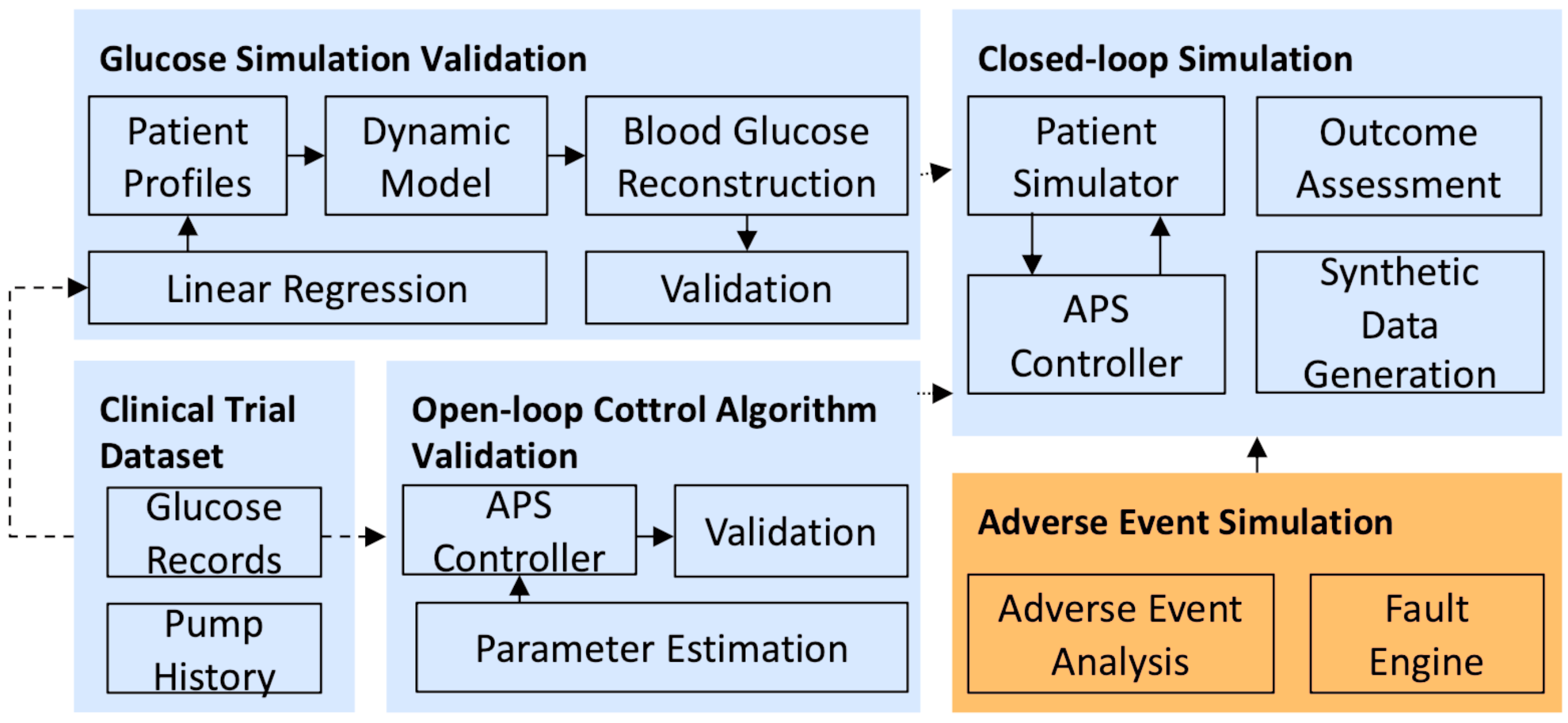}
    \vspace{-2em}
    \caption{Overall Framework for Validation of the APS Testbed.}
    \label{fig:testbed-validation}
\end{figure}

Fig. \ref{fig:testbed-validation} summarizes the overall framework for the validity assessment of the APS testbed. We assess the validity of our testbed using the publicly-available international diabetes closed-loop trial dataset (DCLP3 \cite{brown2019six}). This dataset is collected from a clinical trial of a closed-loop system (t:slim X2 with Control-IQ Technology) \cite{brown2019six} for the six-month treatment of 168 diabetic patients aged 14 to 71 years old, 112 and 56 of which were, respectively, assigned to the closed-loop group (CLC) and the control group that used a sensor-augmented pump (SAP).

To be a valid simulator, the generated data should satisfy the requirements of relevance, completeness, and balance with respect to real-world data \cite{assurance2021Hawkins}.
We ensure relevance and completeness by generating similar patient profiles to those in the real clinical trials (with diverse ages, weights, genders, and medical characteristics) and representative fault/adverse event scenarios that led to FDA recalls.
We measure balance by comparing the percentage of the time the simulated and real trajectories are within the range or contain adverse events.

\subsection{Glucose Simulation}
\label{sec:bgestimate}

To assess the validity of the glucose simulators, we randomly choose five patients' data (each six-month long) from the DCLP3 dataset and compare their BG trajectories during the clinical trial with the simulated BG traces generated using the same insulin inputs from the clinical records at each time step of the simulation. 
At each simulation time step, the insulin rate is set to the corresponding insulin rate at the same time step in the DCLP3 trial dataset. This means that any differences between the BG traces calculated during the simulation and the BG traces measured during the DCLP3 trial are only due to the differences between the simulator's patient model and the actual dynamics of the patient's body. 

However, one challenge in reconstructing each patient's BG trajectory is that some parameters for characterizing the patient profiles in the simulators are not available from the DCLP3 dataset. To solve this problem, we adopt a system identification method to estimate the patient model parameters (patient profiles) from data. We model this problem as the following optimization problem that minimizes the difference between the BG value trajectory reconstructed using the derived parameters and the BG trajectory in the DCLP3 dataset:  
\begin{align}
    param. &= argmin_{param.} \sum_{\mathcal{D}}(BG_{Simulator} - BG_{DCLP3})\\
 s.t. & \forall t \in \mathcal{D} \nonumber \\
    &ID(t)_{Simulator} = ID(t)_{DCLP3} \\
    &BG_{Simulator}(t) = f_{param.}(ID(t-1))
\end{align}

\begin{figure}[t]
    % \vspace{-1.5em}
    \centering
    \includegraphics[width=0.7\columnwidth]{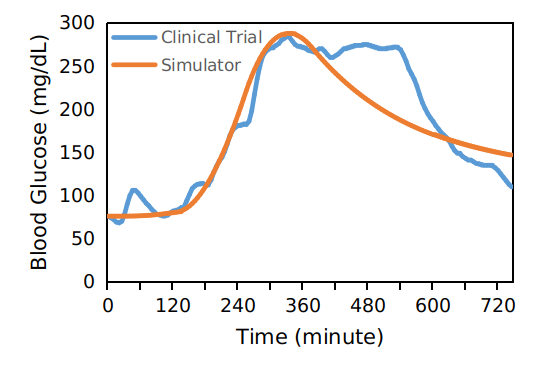}
    \vspace{-2em}
    \caption{BG Trajectories of a Clinical Trial and a Simulation Trace.}
    \label{fig:profileestimation}
\end{figure}

\noindent where $f(\cdot)$ represents the patient model shown in Table \ref{tab:summary-simulator}, $ID(t)$ is the insulin delivery at time step $t$, and $\mathcal{D}$ is the dataset for parameter estimation. We use the linear regression method for parameter estimation and learning based on ten days of data for each patient and the remaining 170 days of data is used to evaluate the validity of the patient simulators.
To reduce the number of parameters that need to be estimated, we also use known metabolic models to directly calculate some unknown parameters from data. For example, the insulin sensitivity factor can be solved by the \textit{1700 rule} \cite{man_uvapadova_2014} using the following equation:
\begin{equation}
    ISF = 1700/TDD
\end{equation}

An example of the BG trajectory using the estimated patient profile and the insulin sequence recorded in the dataset is shown in Fig. \ref{fig:profileestimation}. We see that the reconstructed BG trajectory could approximate the BG values in the clinical trial well in the first 360 minutes but departs from the original trajectory in the last 360 minutes of simulation due to the unpredictable human activity and carbohydrate input. 

\begin{figure}[b]
    \centering
    \includegraphics[width=0.9\columnwidth]{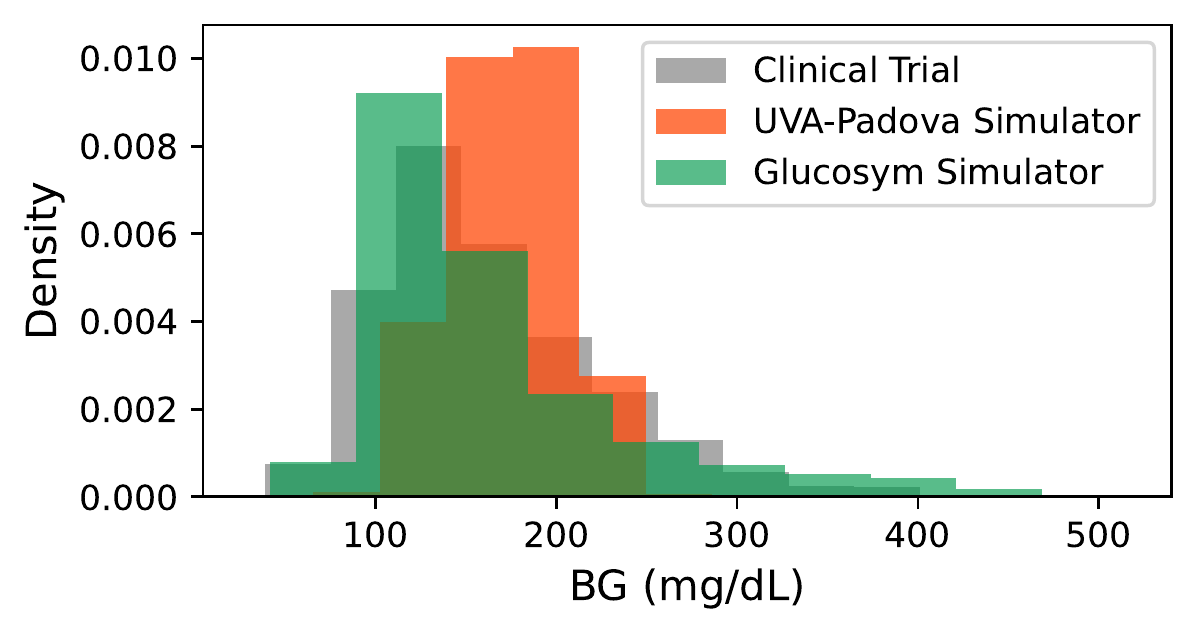}
    \vspace{-1em}
    \caption{Distribution of BG Values: Clinical Trial Data vs. Reconstructed Data by Each Simulator.}
    % \vspace{-1.5em}
    \label{fig:histogram}
\end{figure}
\begin{figure}[t]
    \centering
    % \vspace{-1em}
    \includegraphics[width=0.9\columnwidth]{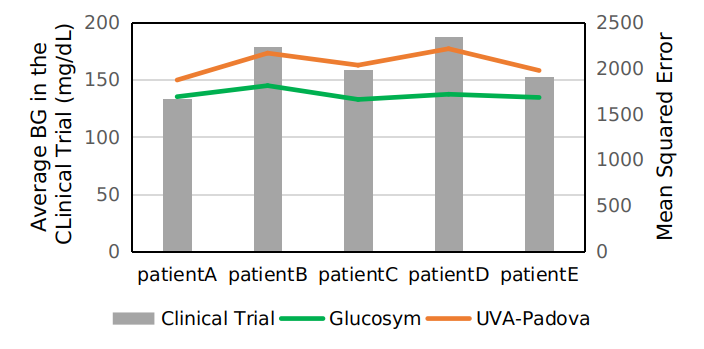}
    \vspace{-1em}
    \caption{Mean Squared Error of BG values between the Clinical Trial and Each Simulator.}
    \vspace{-1.5em}
    \label{fig:bgmse}
\end{figure}

We also present the distribution of BG values reconstructed for all the patient profiles by both the Glucosym simulator and the UVA-Padova simulator in comparison to the baseline BG distribution collected in the DCLP3 dataset in Fig. \ref{fig:histogram}. 
We see that the Glucosym simulator reconstructs the BG distribution that approximates the baseline BG distribution from the clinical trial in both the target range ([70-180] mg/dL) and the high/low BG ranges. On the other hand, the UVA-Padova simulator with the estimated patient profiles generates a BG distribution that is more concentrated between [100-300] mg/dL and does not simulate the extra high/low BG ranges well. We observe similar results when measuring the mean squared error of the BG values estimated for each patient profile using the simulators, as shown in Fig. \ref{fig:bgmse}. This might be because the UVA-Padova uses a more complex dynamic model, and it is more challenging to estimate the patient profiles.

Our preliminary results indicate that with well-tuned patient parameters, the integrated simulators could reproduce similar BG traces from the clinical trial if undergoing the same experimental scenario (i.e., same carbohydrate amount, insulin boluses, and basal pattern, given at the same time). 
The validity of both simulators is also attested to by other researchers who have access to actual patient profiles \cite{man_uvapadova_2014,KanderianT1}, and the fact that the UVA-Padova simulator has been approved by FDA for pre-clinical testing on animals.

% For each real T1DM subject, a virtual subject exists who, if undergoing the same experimental scenario (i.e., same carbohydrate amount, insulin boluses, and basal pattern, given at the same time), behaves similarly from a clinical point of view (i.e., it shows a similar pattern and lies in the same clinically relevant zones [hypo-, eu-, and hyperglycemia]). 
% The distribution of the most important outcome metrics in the simulated traces reproduces those observed experimentally.

% UVA-Padova is approved by FDA and has been assessed with high match to clinic trials in paper

% Glucosym uses real patient profiles

\subsection{APS Control}
\label{sec:controllervalidate}
To assess the validity of the two controllers in the testbed, we feed the BG values from the clinical trial dataset with the same sampling frequency to the different controllers in the testbed, running in an open-loop mode, and compare the output insulin doses calculated by the controllers against the actual pump outputs from the clinical trial.  
We test the validity of both the OpenAPS controller and the Basal-Bolus controller on the closed-loop group (CLC, 112 patients) and the group that used a sensor-augmented pump (SAP group, 56 patients) for six months.

% \usepackage{multirow}
% \usepackage{graphicx}
\begin{table}[b]
\vspace{-1em}
\caption{Insulin Output Comparison Among Each Controller}
\vspace{-1em}
\label{tab:insulin-comparison}
\resizebox{\columnwidth}{!}{%
\begin{tabular}{|c|c|l|c|c|c|}
\hline
\textbf{Metric}                & \textbf{Group} & \textbf{No. Patients} & \textbf{Clinical Trial} & \textbf{OpenAPS}  & \textbf{Basal-Bolus} \\ \hline
\multirow{3}{*}{Avg$\pm$Std} & CLC   & 112         &  0.067 $\pm$  0.061            &  0.067 $\pm$ 0.043        &   0.049 $\pm$ 0.004          \\ \cline{2-6} 
                      & SAP   & 64          &  0.049 $\pm$  0.004   &  0.072 $\pm$  0.045   & 0.049 $\pm$ 0.004            \\ \cline{2-6} 
                      & Avg   & 168         &   0.061 $\pm$ 0.051   &   0.069 $\pm$ 0.044   &    0.049 $\pm$ 0.004         \\ \hline
\multirow{3}{*}{MSE}  & CLC   & 112         & -              & 4.67E-03 &  4.01E-03           \\ \cline{2-6} 
                      & SAP   & 64          & -              & 2.51E-03         & 4.74E-06            \\ \cline{2-6} 
                      & Avg   & 168         & -              & 3.97E-03         & 2.70E-03    \\ \hline
\end{tabular}%
}
\end{table}

As shown in Table \ref{tab:insulin-comparison}, we calculate the Mean Squared Error (MSE) between the simulated controller and the actual pump outputs. We see that the Basal-Bolus produced a much smaller MSE in the SAP group than the OpenAPS controller and maintained a lower MSE in the CLC group. The OpenAPS controller uses a different control algorithm from the pump used in the clinical trial.

\begin{figure}[b]
    \vspace{-1.5em}
    \centering
    \includegraphics[width=\columnwidth]{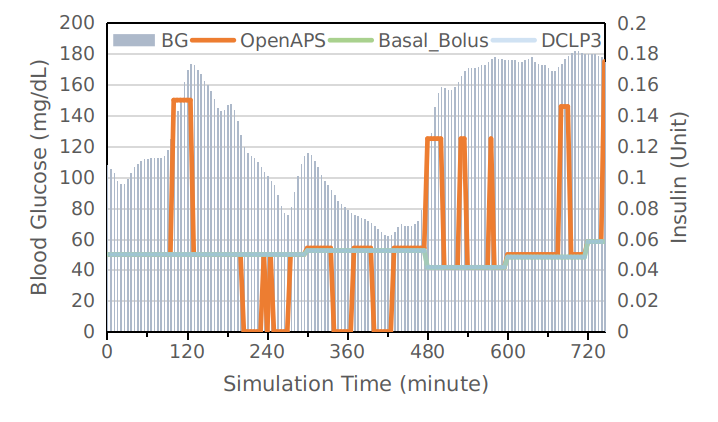}
    \vspace{-2.5em}
    \caption{An Example of the Insulin Outputs among Each Controller Given the Same Glucose Readings.}
    % \vspace{-1.5em}
    \label{fig:patient3insulin}
\end{figure}

An example of the insulin output comparison of both the OpenAPS controller and the Basal-Bolus controller to the control algorithm used by the pump in the clinical trial is shown in Fig. \ref{fig:patient3insulin}. We observe that the Basal-Bolus controller can well reproduce the control actions made by the insulin pump used in the clinical trial, demonstrating the validity of the integrated controller in matching actual insulin pump control actions. On the other hand, the OpenAPS controller makes different decisions when the predicted BG is going outside of the target range of 70 t o180 mg/dL or a risk of hyperglycemia or hypoglycemia is anticipated. It should be noted that the OpenAPS controller uses the exact control software used by actual diabetic patients. This is consistent with the observation from previous studies that showed OpenAPS has better performance than some of the existing commercial pumps \cite{melmer2019glycaemic,lewis2016real}.

% \textbf{how to evaluate}
% percentage of  
% the time to keep BG within the normal range or target range;
% the time or values in Hypoglycemia or Hyperglycemia.

% Number of events 
% Hypoglycemia or Hyperglycemia.

% \usepackage{graphicx}
\begin{table*}[tp!]
\small
\caption{Comparison of the Outcomes between Closed-loop Simulation and the Clinical Trial}
\vspace{-1em}
\label{tab:outcome}
% \resizebox{\columnwidth}{!}
{
\begin{threeparttable}
\begin{tabular}{|l|l|l|l|}
\hline
\textbf{Outcome}                             & \textbf{Clinical Trial} & \textbf{Closed-loop1$^{1}$} & \textbf{Closed-loop2$^{2}$} \\ [0.5ex] \hline
Pct. of time with BG in target range of 70 to 180 mg/dL        & 66.18±25.56    & 93.49±10.67   & 79.46 ± 16.68 \\ [0.5ex]\hline
Pct. of time with BG\textgreater{}180 mgl/dL & 32.07±25.83    & 3.95±7.34    & 20.16 ± 16.08 \\ [0.5ex]\hline
Pct. of time with BG\textless{}70 mgl/dL     & 1.75±4.67      & 2.56±7.06    & 0.05 ± 0.24   \\ [0.5ex]\hline
Pct. of time with BG\textless{}54 mgl/dL     & 0.33±1.67      & 0.12±1.63    & 0.02 ± 0.14  \\ [0.5ex]\hline
\end{tabular}

\begin{tablenotes}
 \item[1] Glucosym simulator with OpenAPS control software.
 \item[2] UVA-Padova simulator with Basal-Bolus controller.
\end{tablenotes}
\end{threeparttable}
}
\vspace{-1em}
\end{table*}

\subsection{Closed-loop Simulation}
Finally, we run the controllers and simulators together in a closed-loop mode to assess their performance when automatically regulating the blood glucose of diabetic patients. For this kind of assessment, we cannot compare the BG readings or the insulin outputs step by step between the simulation and the clinical trial, as a differing control action changes the subsequent BG values. Instead, we adopt a metric that evaluates the primary and secondary outcomes in diabetes treatment \cite{brown2019six} by measuring the percentage of time that the BG value is inside or outside the target range of 70 to 180 mg/dL. 

We randomly select five patients to estimate their profiles and run both simulators with OpenAPS and Basal-Bolus control software, respectively, in a closed-loop using the patient profiles and other required parameters estimated in Section \ref{sec:bgestimate}-\ref{sec:controllervalidate}.

% We selected the data of 5 patients from the DCLP3 dataset that contains BG value outside target range (70-180 $mg/dL$) and estimate the patient profiles required by GLucosym simulator and UVA-Padova simulator using system identification method. We run both simulators with OpenAPS and Basal-Bolus control software respectively in a closed loop.

% When the T1DM simulation was run with the BB controller, the same estimated patient profile calculated in Section 4.1 was used. 

% Our closed-loop simulation can not only very well reproduce the clinical trials but also reduce adverse events, which offers a quantitative evaluation of the percentage of time with BG value in different range and can help improve the pump algorithm. 

Experiment results in Table \ref{tab:outcome} show that both simulated closed-loop APS maintain a higher percentage of time with BG inside the target range than the baseline control system used in the clinical trial,
though the closed-loop system with UVA-Padova simulator and Basal-Bolus controller has a closer outcome with the clinical trial as they use a similar control algorithm. 
% moved up because both these paragraphs reference table 9
We also observe that the closed-loop system with the Glucosym simulator and OpenAPS controller software keeps the BG in the target range for 93.49\% of the time on average, demonstrating the advance of this PID-based control algorithm in diabetes treatment over the regular insulin therapy with Basal-Bolus control algorithm.

\begin{figure}[b]
    \centering
    \vspace{-2em}
    \includegraphics[width=0.95\columnwidth]{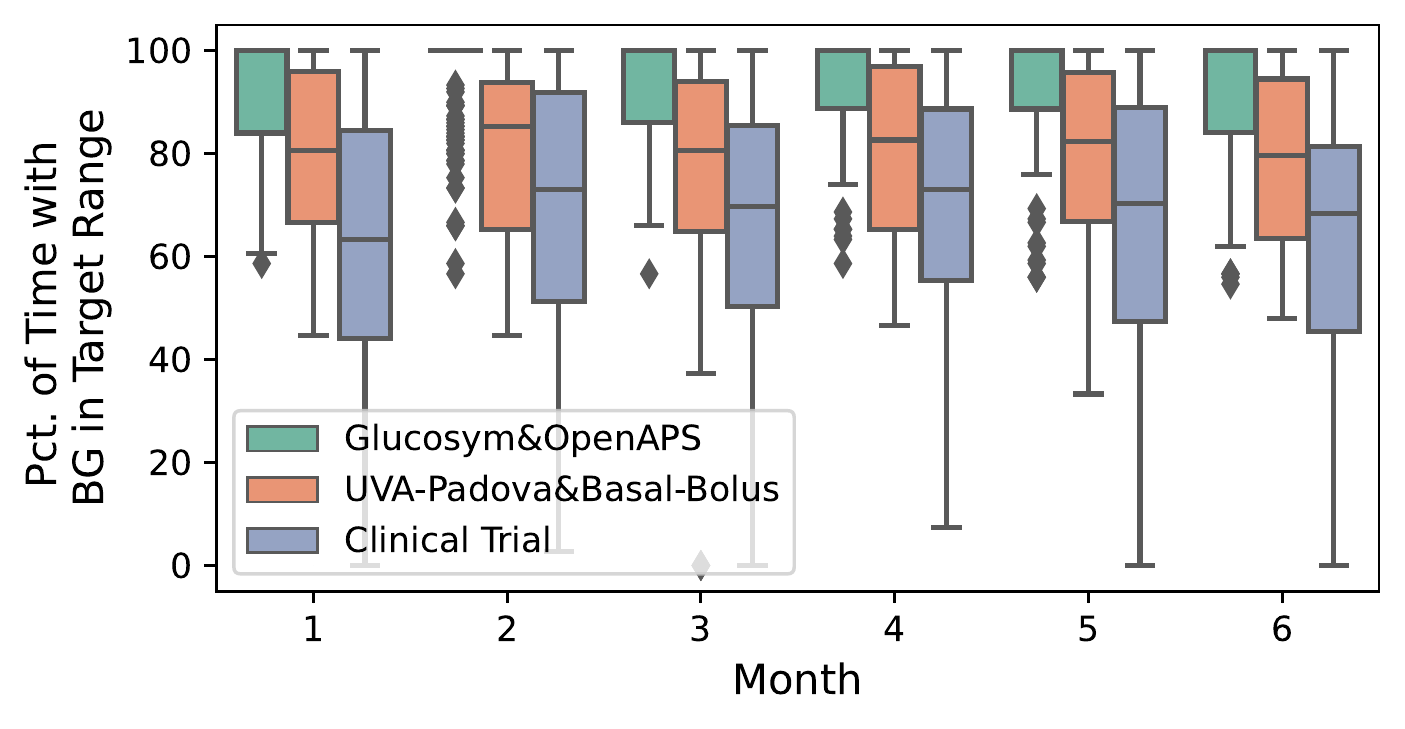}
    \vspace{-1em}
    \caption{Percentage of Time with BG in the Target Range of 70 to 180 mg/dL for Clinical Train and Two Closed-loop APS.}
    % \vspace{-1em}
    \label{fig:targetRangebyMonth}
\end{figure}

The primary outcomes of each closed-loop APS and the actual control system in the clinical trial across six months are shown in Fig. \ref{fig:targetRangebyMonth}. We see that the mean percentage of time with glucose values within the target range remained at a similar level during the six months in the clinical trial and both closed-loop simulations. 

\begin{figure*}[tp]
    \centering
    % \vspace{-1em}
    \begin{minipage}{0.95\columnwidth}
	\scriptsize \centering
    \includegraphics[width=\columnwidth]{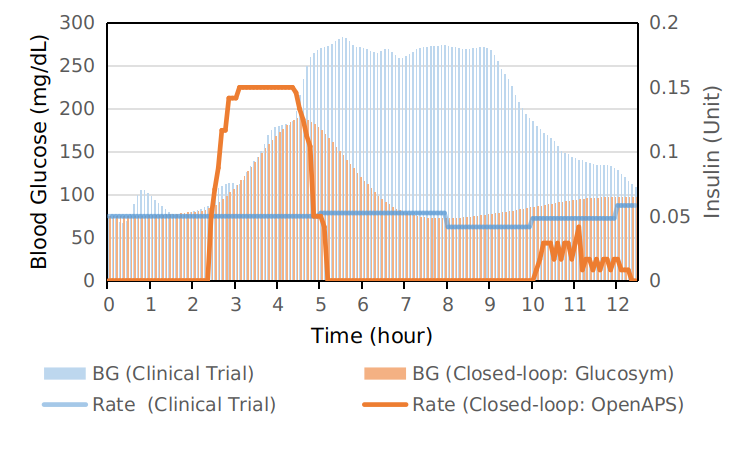}
    
    % (a)  
    \end{minipage}
    %\hfill
    \begin{minipage}{0.95\columnwidth}
	\scriptsize \centering
    \includegraphics[width=\columnwidth]{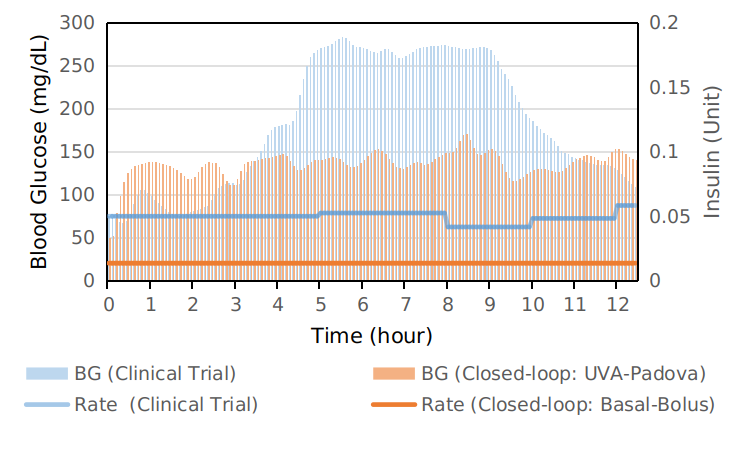}
    
    % (b)  
    \end{minipage}
    
    \vspace{-2em}
    \caption{Comparison of BG and Insulin Rate between a Clinical Trial and the Closed-loop Simulation.}
    \label{fig:closeloopexample}
    \vspace{-0.5em}
\end{figure*}

The closed-loop simulation offers a platform to evaluate or improve different pump algorithms with various patient profiles. For example, Fig. \ref{fig:closeloopexample} shows the different decisions made by each control algorithm at each time step during 12.5 hours of treatment/simulation. We see that the actual pump from the clinical trial used a fixed basal rate and thus failed to keep the BG within the target range, resulting in adverse hyperglycemia. In comparison, the OpenAPS kept the BG value safe by increasing the insulin infusion when the BG is predicted to increase quickly and keeping a low insulin dose when the insulin on board is still at a high level after a large amount of previous insulin injection. Similarly, the Basal-Bolus controller with the UVA-Padova simulator issued a higher basal rate to avoid the BG value increasing too fast. 
Through such simulation and comparison, the proposed testbed can help to improve different control algorithms used in commercial insulin pumps and reduce patient harm or complaints.

\subsection{Adverse Event Simulation}

From Table \ref{tab:recall} and Table \ref{tab:outcome}, we see that adverse events naturally happen during clinical trials or home treatment of diabetic patients with APS due to device malfunctions or control software defects. However, it is expensive or risky for the manufacturers to test and improve the control algorithms through experiments on actual patients within realistic environments. The proposed closed-loop simulation offers an alternative way to evaluate the effectiveness of different control algorithms with actual patient profiles. However, the rate of adverse events in the closed-loop simulation is too low to evaluate the resilience of the target control algorithm comprehensively. 
Therefore, the adverse event simulator proposed in Section \ref{sec:adversesimulator} was used to simulate the following scenarios: hyperglycemic adverse events (diabetic ketoacidosis), hypoglycemic events, malfunction of the device, and injury of patients.

\begin{figure}[b!]
    \vspace{-1em} 
    \centering
    \begin{minipage}{0.95\columnwidth}
	\scriptsize \centering
    \includegraphics[width=\columnwidth]{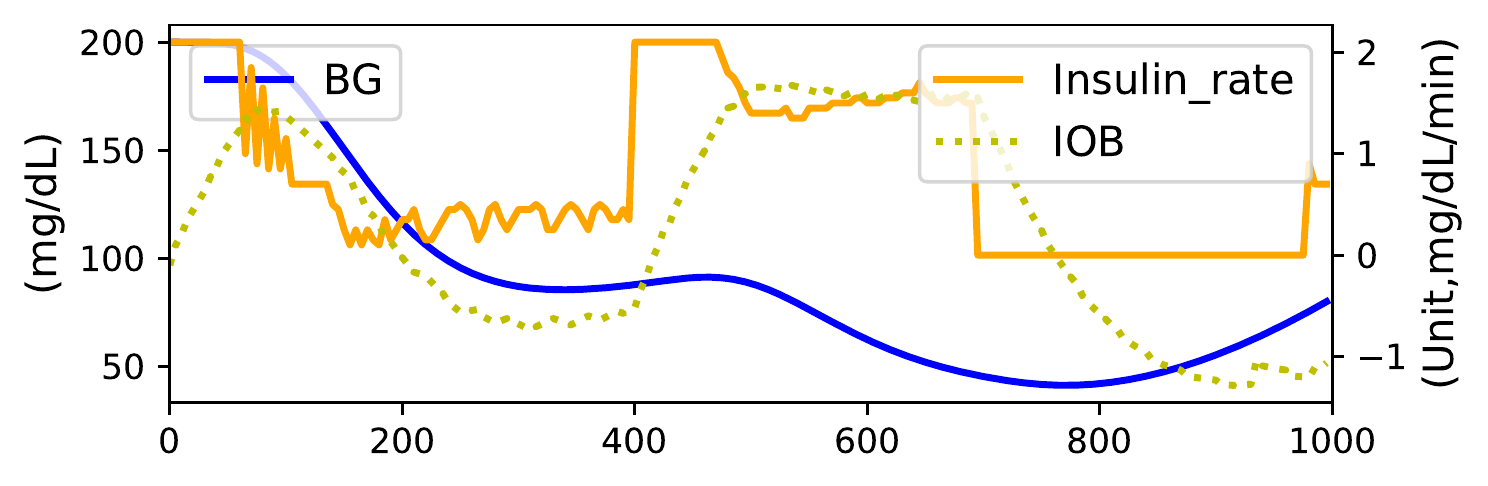}
    
    % (a)  
    \end{minipage}
    %\hfill
    \begin{minipage}{0.95\columnwidth}
	\scriptsize \centering
    \includegraphics[width=\columnwidth]{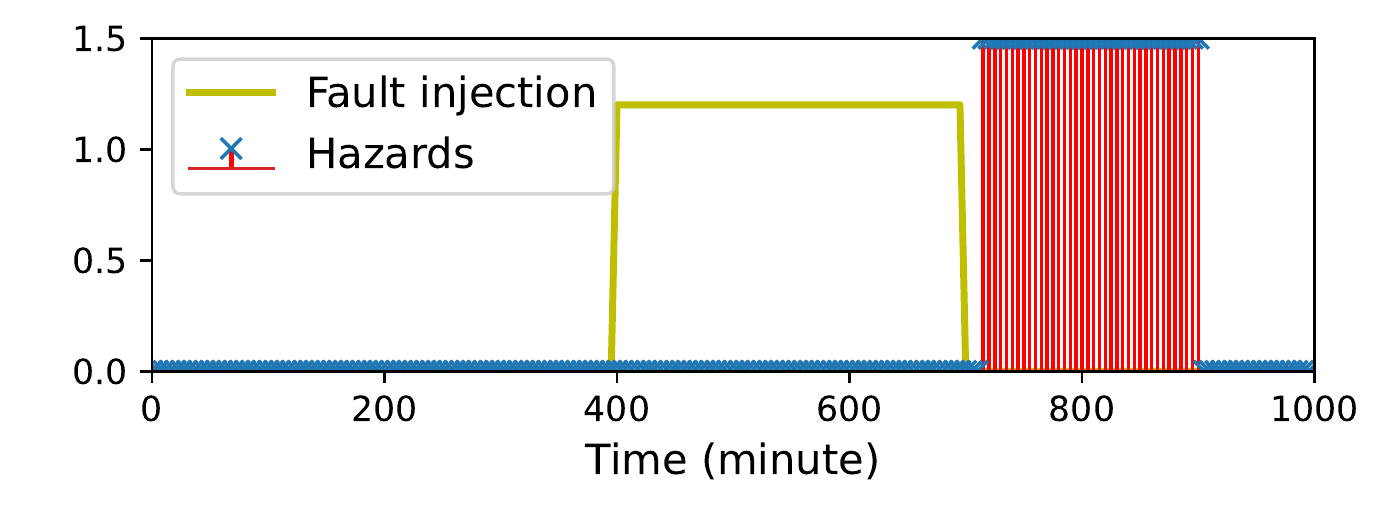}
    
    % (b)  
    \end{minipage}
    
    \vspace{-1em}
    \caption{An Example of Simulated Hypoglycemic Adverse Event due to Fault Injection.}
    \label{fig:traj}
    % \vspace{-1.5em}
\end{figure}

We run 882 simulations (14 fault scenarios, as \textit{Add/Sub} includes multiple sub-scenarios, times 9 random start times and durations times 7 initial BG values), and each simulation includes 12.5 hours of sensor data and insulin outputs after a meal with different carbohydrate inputs. 
% Show example figures (corresponding to Fig. 5 which is fault free) with different types of faults injected and mention that this helps with assessing the robustness of different algorithms against faults in the sensor and actuator or control algorithm itself.
An example of the BG trajectory with a fault injected starting at 400 minutes to simulate a CGM sensor reading malfunction is shown in Fig. \ref{fig:traj}. We see that the controller increased the insulin injection significantly based on the erroneous CGM readings while the BG value was not high, which further decreased the BG value under 50 mg/dL and resulted in a severe hypoglycemic event (marked by the red region in Fig. \ref{fig:traj}).   

\begin{figure}[b!]
    \centering
    \vspace{-2em}
    \includegraphics[width=\columnwidth]{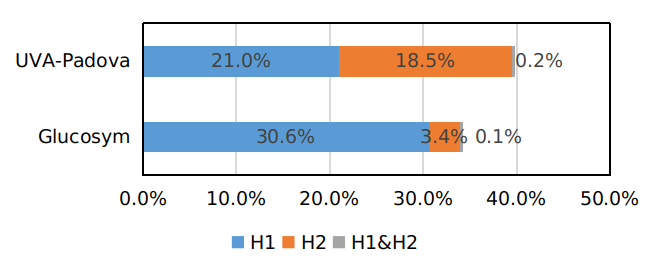}
    \vspace{-2em}
    \caption{Success Rate of the Fault Injection Experiments in Simulating Low BG Hazard (H1) and High BG Hazard (H2), which may Result in Hypoglycemic or Hyperglycemic Adverse Event. }
    % \vspace{-1.5em}
    \label{fig:hazardrate}
\end{figure}

Our simulation generates two and a half years of data for 20 diabetes patients with different types of adverse events. The percentage of adverse events in the Glucosym simulator and UVA-Padova simulator are 33.9\% and 39.3\%, respectively (see Fig. \ref{fig:hazardrate}). The generated synthetic dataset is available on Github [\url{https://github.com/UVA-DSA/APS_TestBed}].

\section{Related Work}

\textbf{MCPS Testbeds:}
% The need for high fidelity testbeds is ubiquitous across CPS. Development of CPS testbeds began with traditional CPS, such as smart grids \cite{zhang_multifunctional_2017, stanovich_development_2013, poudel_real-time_2017}, but has since been extended to the medical domain, a safety-critical subset of CPS. 
Testbeds are used in place of time- and resource-intensive clinical trials, so development has gravitated towards systems that affect the most critical organs. 
For instance, heart testbeds have been constructed for pacemaker validation \cite{zhihao_jiang_using_2010} and cardiovascular interventions \cite{vrooijink_beating_2018}. Robotic surgery testbeds have also been made for MRI-guided biopsy \cite{mendoza_testbed_2019}, endovascular surgery \cite{karstensen_facile_2021}, and reconstructive surgeries in the hand \cite{tigue_simulating_2020}. In-silico trials of an insulin control algorithm are developed recently to facilitate research on APS \cite{schmitzer2022efficient,schmitzer2022efficient}. 
% Pulse \cite{bray_pulse_2019}, a software engine, provides a comprehensive testbed for the cardiovascular system using digital circuit analogs. Medical device plug and play (MDPnP) validation protocols \cite{wu_treatment_2014, arney_toward_2010} can be utilized to verify a treatment before it is ever implemented, and was shown to be effective for cardiac arrest and PCA infusion pump case studies.
To the best of our knowledge, this paper is the first to develop an open-source closed-loop testbed for APS with real-world controllers, physical simulators of an extended set of patient profiles, and a realistic adverse event simulator.

% \vspace{-1em}

% APS work
\textbf{Glucose Simulators:}
The use of simulators is vital in the development of APS. A glucose minimum model, a simple mathematical model for glucose levels, was first proposed in 1970 \cite{Insel1974modeling}.
% Simulators
% UVA-Padova - approved by FDA for clinical trials (check extent)
% Cambridge - released around the same time as the UVA-Padova sim, increased the number of virtual patients
% OSHU - the glucoregulatory model used is similar to the Cambridge glucoregulatory model, except the insulin kinetics model is different.
% Glucosym - open source simulator designed for ease of testing
The UVA-Padova simulator \cite{Kovatchev2009insilico} was the first APS simulator to be approved by the FDA in 2008 as a substitute for animal testing. A second simulation developed by a group at Cambridge University was released soon after in 2010, specifically targeted toward closed-loop APS simulation and virtual patient modeling \cite{wilinska_simulation_2010}. The UVA-Padova was updated in 2014 \cite{man_uvapadova_2014} to improve the glucose kinetics model during hypoglycemia as well as incorporate glucagon kinetics and was accepted by the FDA as a substitute for certain preclinical trials. Glucosym, an open-source APS simulator, was released in 2017 to widen the availability of closed-loop APS simulation and testing \cite{Glucosym}. In 2019, a group at the Oregon Health and Science University (OHSU) published an APS simulator based on a similar glucoregulatory model as the Cambridge simulator but with different insulin kinetics \cite{resalat_statistical_2019}. 
Our work differentiates from these previous works by integrating two advanced control software with the state-of-the-art simulators into a closed-loop APS testbed and proposing an optimization method to estimate real patient profiles for the closed-loop simulation.

% Safety work - split into attacks and defense on CPS
\textbf{Safety Evaluation of APS: }
Many previous works have also focused on evaluating the safety of APS control software, such as safety and effectiveness evaluation of insulin pump therapy in children and adolescents with Type 1 diabetes
\cite{plotnick2003safety,alotaibi2020efficacy}, safety and efficacy review of commercial and emerging hybrid closed-loop systems \cite{fuchs2020closed,hirsch2008sensor}, generic safety requirements for developing safe insulin pump software \cite{zhang2011generic} and insulin pump software certification \cite{chen2013insulin}, or safety evaluation of do-it-yourself APS \cite{toffanin2020silico}. However, most of these works relied on high-risk clinical tests or were not able to assess the resilience of tested insulin pumps against adverse events.
In this paper, we integrate an adverse simulator into the closed-loop APS testbed, which could help with the evaluation of different APS control algorithms and safety mechanisms in preventing adverse events while avoiding actual harm to the patients.

% To address these concerns, proposed authentication protocol for these devices leverages artificial intelligence to scan biometric data in an effort to resolve security and privacy issues \cite{qi_security_2021}. If an intruder does manage to gain access to the network, detection algorithms that use behavior-rule specification-based techniques can be used to alarm the patient \cite{mitchell_behavior_2015}. Neural networks can also be utilized to detect anomalies in the commands issued by an APS controller for passive noise and attack protection \cite{zhou_robustness_2022}. Threat and trust modelling frameworks were developed to make security protocols more robust for MCPS \cite{almohri_threat_2017}. Additionally, the concept of context-awareness has been incorporated into MCPS to detect human error and others problems and mitigate harmful impacts \cite{zhou2021data, li_towards_2015}.

\section{Conclusion}
% \vspace{-0.5em}
\label{sec:conclusion}
% main points: testbed is a valid tool that can be used in future work to demonstrate the effectiveness of the control algorithm and/or safety features for APS

Using two state-of-the-art glucose simulators, we develop a testbed for evaluating the performance of the control algorithms and safety features in APS. We assess the validity of the simulator by reverse-engineering the profiles of patients in a real clinical trial and demonstrating that the BG traces generated by each simulator are functionally the same as the BG traces from the trial. We also show the testbed's utility for closed-loop simulation by implementing two control algorithms to regulate the glucose levels of virtual patients. To push the APS to its limit, we embed a novel fault injection engine based on real FDA recalls into the testbed so performance can be evaluated in even the most hostile scenarios. 

%This work has shown that the proposed testbed can be used in future work to demonstrate the effectiveness of new, more powerful control algorithms and safety features in APS. 
As research turns toward adopting more advanced data-driven methods like machine learning for the design of control algorithms, the proposed testbed and other \textit{in silico} testing strategies will be essential for both final product evaluation and sourcing large quantities of high-quality data. 
This testbed can also be used to further develop personalized treatments by tailoring control algorithms to individual or similar patient profiles and to help diabetic patients understand their treatments by modeling their physiological dynamics.
In future work, our testbed could be improved by developing more accurate estimation methods for patient profiles and incorporating meal and activity models and simulators.
%We hope this testbed can facilitate the development of safe and robust APS that will improve the lives of those living with Type 1 diabetes.

% \vspace{-1.25em}
% \subsection*{Code Availability}
% \vspace{-0.75em}
%The code for the closed-loop APS system and the safety monitor is available at: [Removed for anonymous submission]

% \url{ https://github.com/UVA-DSA/openAPS_GlucoSym_closed_loop_system}.
% \vspace{-1em}
% \textit{https://github.com/UVA-DSA/openAPS\_GlucoSym\_closed\_loop\_system.git}

% \vspace{-1em}
\section*{Acknowledgment}
This work was partially supported by the Commonwealth of Virginia under Grant CoVA CCI: C-Q122-WM-02 and by the National Science Foundation (NSF) under Grant No. 2146295.

\bibliographystyle{ACM-Reference-Format}
% \bibliography{main}
\bibliography{main}

\end{document}